\documentclass{aa}  

\usepackage{graphicx}
\usepackage{txfonts}
\usepackage{natbib}
\usepackage[colorlinks, allcolors=blue]{hyperref}
\usepackage{xcolor}

\begin{document}

\title{The chemical DNA of the Magellanic Clouds}
\subtitle{V. $R$-process dominates neutron capture elements production \\ in the oldest SMC stars
\thanks{Based on observations collected at the ESO-VLT under the program 111.24Z9 and at Las Campanas Observatory under the programs CN2023A-37, CN2023B-37.}}

\author{L. Santarelli\inst{1,2}
\and
M. Palla\inst{1,2,3}
\and
A. Mucciarelli\inst{1,2}
\and
L. Monaco\inst{4,5}
\and
D. A. Alvarez Garay\inst{3}
\and
D. Romano\inst{2} 
\and
C. Lardo\inst{1,2}
}

\institute{Dipartimento di Fisica e Astronomia “Augusto Righi”, Alma Mater
Studiorum, Università di Bologna, Via Gobetti 93/2, 40129 Bologna, Italy
\and
INAF – Osservatorio di Astrofisica e Scienza dello Spazio, Via Gobetti 93/3, 40129 Bologna, Italy
\and
INAF - Osservatorio Astrofisico di Arcetri, Largo E. Fermi 5, 50125, Firenze, Italy
\and
Universidad Andres Bello, Facultad de Ciencias Exactas, Departamento de F{\'i}sica y Astronom{\'i}a - Instituto de Astrof{\'i}sica, Autopista Concepci\'on-Talcahuano 7100, Talcahuano, Chile
\and
INAF-OATs, Via G.B.Tiepolo 11, Trieste, I 34143, Italy\\
}

\date{Received 29 October 2025; accepted 12 December 2025}

\abstract{We present the chemical abundances of Fe, $\alpha$- and neutron-capture elements in 12 metal-poor Small Magellanic Cloud (SMC) giant stars, observed with the high-resolution spectrographs UVES/VLT and MIKE/Magellan. These stars have [Fe/H] between $-2.3$ and $-1.4$ dex, 10 of them with [Fe/H]$<-1.8$ dex. According to theoretical age-metallicity relations for this galaxy, these stars formed in the first Gyr of life of the SMC and represent the oldest SMC stars known so far. [$\alpha$/Fe] abundance ratios are enhanced but at a lower level than MW metal-poor stars, as expected according to the slow star formation rate of the SMC. The sample exhibits a large star-to-star scatter in all the neutron-capture elements. The two $r$-process elements measured in this work (Eu and Sm) have abundance ratios from solar up to +1 dex, three of them with [Eu/Fe]$>+0.7$ dex and labelled as $r$-II stars. This [$r$/Fe] distribution indicates that the $r$-process in the SMC can be extremely efficient but is still largely affected by the stochastic nature of the main sites of production and the inefficient gas mixing in the early SMC evolution. A similar scatter is observable also for the $s$-process elements (Y, Ba, La, Ce, Nd), with the stars richest in Eu also being rich in these $s$-elements. Also, all the stars exhibit subsolar [$s$/Eu] abundance ratios. At the metallicities of these stars, the production of neutron-capture elements is driven by $r$-process, because the low-mass AGB stars have not yet evolved and left their $s$-process signature in the interstellar medium. We also present stochastic chemical evolution models tailored for the SMC that confirm this scenario.}


   \maketitle

\section{Introduction}

    Metal-poor stars provide unique insights into the earliest phases of chemical enrichment in the Universe \citep[see][and references therein]{Bonifacio25}. They are the oldest stars we can reach, preserving in their photosphere the nucleosynthetic signatures of the first generations of massive stars and supernovae (SNe). In particular, the abundances of elements heavier than the iron-peak offer a powerful diagnostic of the astrophysical sites and the timescales of the neutron-capture processes, in which these elements formed from seed nuclei. Neutron-capture processes are generally subdivided in slow ($s$-) or rapid ($r$-) processes\footnote{Other, secondary neutron-capture processes have also been identified, the most relevant being the intermediate process \citep{Cowan77}.}, depending on the timescale of neutron captures with respect to that of $\beta$-decays, an interplay that in turn depends on the density of the physical source of neutrons and on the consequently available neutron flux \citep{Burbidge57, Cameron1957, Clayton68}. 

    Slow neutron-capture process occurs at intermediate neutron densities ($\sim10^{7-12}$ neutrons~$\cdot$ cm$^{-3}$), when each neutron capture is followed by a $\beta$-decay \citep{Burbidge57, Kappeler11}. Low-mass asymptotic giant branch (AGB) stars are the main sites of production of $s$-process elements \citep[e.g.,][]{Ulrich73,Iben83,Straniero06,Karakas14}. However, a non negligible contribution is also provided by rotating massive stars \citep{Limongi18}, especially for first peak of $s$-process elements (e.g. Sr, Y, Zr) and marginally for second peak of $s$-process elements (e.g. Ba, La, Ce, Nd).

    Rapid neutron-capture process takes place when extreme neutron fluxes ($>10^{20}$ neutrons~$\cdot$ cm$^{-3}$) determine multiple neutron captures on a timescale much shorter than the $\beta$-decays, that occur only afterwards and bring the nuclei back towards the valley of stability \citep{Cowan21}. In the case of $r$-process, the main sites of production are still debated. The current view is that two competitive sources of production, a prompt and a delayed source, contribute on different timescales. The prompt source should include the contribution by peculiar core-collapse SNe (CC-SNe), like magneto-rotationally driven SNe (MRD-SNe)/collapsars \citep{Nishimura17,Siegel19}, proto-magnetars \citep{Nishimura15} and common-envelope jets SNe \citep{Grichener19}. The delayed source, occurring on timescales larger than that of the CC-SNe, is identified with merging of compact objects \citep[neutron star-neutron star or neutron star-black hole; hereafter simply referred to as NSM,][]{Lattimer74,Argast04}. Both sources of $r$-process need to be included in the chemical evolution models for the Milky Way (MW) in order to reproduce the observed patterns for $r$-process elements \citep[e.g.][]{Molero21}.
    
    MW metal-poor stars exhibit presence of neutron-capture elements formed, mostly, by the $r$-process or weak $s$-process, coupled with significant star-to-star scatter that likely reflect the stochastic nature of the rare events producing these elements (for $r$-process in particular) and inhomogeneous interstellar medium (ISM) mixing in the early Galaxy. On the other hand, the study of neutron-capture elements in metal-poor stars of external galaxies is still an unexplored field. Stars with [Fe/H] between $-3.5$ and $-2.5$ dex have been discovered in irregular, dwarf and ultra-faint dwarf galaxies \citep{Tafelmeyer10, Frebel16, Simon19}, expanding our comprehension of the early chemical enrichment of the Universe and opening a new field of study in astrophysics. However, despite their proximity, the investigation of metal-poor stars in the Large and Small Magellanic Cloud (LMC and SMC, respectively) has received limited (and only recent) attention \citep[see e.g.][]{Reggiani21,Chiti24,Oh24}.

    Unlike the LMC, where the oldest stellar populations can be investigated through a populous family of 15 old globular clusters (GCs) with ages, masses, and properties similar to those of the Galactic halo clusters \citep{Johnson06, mu10, mu21b}, the earliest generations in the SMC cannot be studied with these diagnostics. In fact, the oldest SMC GC, named NGC~121, has an age of 10.5$\pm$0.5 Gyr \citep{Glatt08}, with [Fe/H]$=-1.18\pm0.02$ dex and almost solar [$\alpha$/Fe] \citep[][hereafter Paper II]{mu23b}, indicating that the enrichment at this age is already dominated by SNe~Ia. According to the age-metallicity relation by \citet{Pagel98}, in order to study SMC stars older than 12 Gyr we need to identify field stars with [Fe/H]$\lesssim-1.6$, that represent a small fraction of the total stellar content in this galaxy \citep{Carrera08, Nidever20, mu23a}. 

    Chemical abundances for SMC field stars with [Fe/H]$\lesssim-1.6$ dex, likely corresponding to the oldest populations in this galaxy, are still restricted to small samples with incomplete information. The APOGEE survey has observed a large number of red giant branch (RGB) stars across the entire extensions of the Clouds, providing abundance for Fe, $\alpha$ and some key iron-peak elements \citep{Nidever20, Hasselquist21}. However, the sample of SMC stars selected by \citet{Hasselquist21} includes 55 stars with [Fe/H]$<-1.6$ dex and among them only nine with [Fe/H]$<-2.0$ dex. Also, the H-band spectroscopy is not suitable to study neutron-capture elements \citep[see e.g.][]{Manea25}, especially those produced through $r$-process, like Eu. As explained by \citet{Smith21}, only few lines of the $s$-process elements Ce II and Nd II were detectable in the H-band, while the only $r$-process dominated feature is a weak absorption line of Yb II.

    \citet[][R21 afterwards]{Reggiani21} performed a high-resolution spectroscopic observations of nine LMC and four SMC candidate metal-poor stars. The four SMC giants have metallicities $-2.6\lesssim$[Fe/H]$\lesssim-2.0$, exhibiting chemical patterns analogue to those observed in MW stars of similar [Fe/H]. However, the only two with available Eu abundances in their SMC sample exhibit a highly significant enhancement in this $r$-process element, with [Eu/Fe] $=+0.85$ and $+1.00$ dex; the offset is even more remarkable for LMC stars and for the whole MCs sample. This larger abundance of Eu at lower metallicities was interpreted as the result of the Clouds isolated chemical evolution and long history of gas accretion from the cosmic web, combined with $r$-process nucleosynthesis on a timescale longer than that of CC-SNe but shorter or comparable to that of SNe Ia.

    In the first paper of this series, \citet[][hereafter Paper I]{mu23a}, we presented the chemical composition of 206 RGB SMC stars observed with FLAMES-GIRAFFE optical spectrograph at the Very Large Telescope and some tens of metal-poor ([Fe/H]$\lesssim-1.6$ dex) stars were identified. For these metal-poor stars, a significant abundance scatter in the $s$-process element Ba was found. \citet[][hereafter Paper IV]{Anoardo25} performed a spectroscopic follow-up of the sample of Paper I devoted at measuring [Eu/Fe], not measurable with the spectral dataset used in Paper I. The SMC stars are characterised by a strong enhancement of [Eu/Fe] at any metallicity, with a clear decreasing run by increasing [Fe/H], reflecting the contribution by SNe Ia. The few metal-poor stars analysed in Paper IV suggest a significant star-to-star scatter in $r$-process element Eu.

    In the light of the above findings, this work aims at a deeper delving into the abundances of metal-poor stars in the SMC. Thanks to new follow-up observations of SMC metal-poor stars by means of high-resolution spectrographs, access to an extended (from oxygen to europium) and dense (e.g. 7 neutron-capture elements analysed) chemical inventory is possible, allowing a detailed discussion on the origin and the correlations between different neutron-capture elements in this galaxy.\\

    The work is organised as follows: Sect.~\ref{dataset} presents the spectroscopic dataset; Sect.~\ref{analysis} describes the spectral analysis; Sect.~\ref{iron} and ~\ref{abundances} present the derived chemical abundances; finally, in Sect.~\ref{ss:rproc_sproc} and ~\ref{ss:stoc_CEM} we provide an interpretation of the measured abundance patterns in the framework of the chemical enrichment history of the SMC, using a new stochastic chemical evolution model. Finally, in Sect.~\ref{conclusions} we draw our conclusions.

\section{Spectroscopic dataset} \label{dataset}
    
    The dataset discussed here consists of optical spectra for 12 metal-poor RGB stars of the SMC, whose IDs and main information are summarised in Table \ref{id_coord}. Seven stars are selected from the sample of SMC field stars by Paper I. They were observed with the Ultraviolet and Visual Echelle Spectrograph \citep[UVES,][]{Dekker00} at the Very Large Telescope (ID Program: 111.24Z9, PI: Mucciarelli),  adopting a 1" slit (R$\sim$40000). With the dichroic mode both the Blue Arm CD2 390 and the Red Arm CD3 580 settings were observed simultaneously, providing for each target spectra in ranges 3260-4520 \AA, 4780-5760 \AA, and 5830-6800 \AA. Exposure times are 5$\times$3000s for UVES-1, UVES-6, 3$\times$3000s for UVES-2, UVES-3 and UVES-5, and 4$\times$3000s for UVES-4 and UVES-7. The SNR per pixel of the combined spectra is $\sim50$ at 5300 \AA\;and $\sim60$ at 6300 \AA. 

    \begin{table*}
        \caption{Main information about the spectroscopic targets.}
        \label{id_coord}
        \centering
        \begin{tabular}{c c c c c c c c c}
        \hline\hline
        ID & Gaia DR3 ID & Paper I/ & RA & Dec & $G$ & $K_s$ & $E(B-V)$ & R21 \\
        & & APOGEE ID & & & & & &\\
        \hline
        & & & [degree] & [degree] & [mag] & [mag] & [mag] \\
        \hline
        UVES-1 & 4689859031717528576 & FLD-121\textunderscore100330 &  6.599776 & -71.481933 & 16.82 & 14.30 & 0.028 & *\\
        UVES-2 & 4689864941592504064 & FLD-121\textunderscore100514 &  6.665003 & -71.369523 & 16.14 & 13.46 & 0.028 & *\\
        UVES-3 & 4689861093301638272 & FLD-121\textunderscore100683 &  6.906465 & -71.419883 & 16.00 & 13.24 & 0.028 & *\\
        UVES-4 & 4689844153950876928 & FLD-121\textunderscore100767 &  6.516431 & -71.672640 & 16.44 & 13.78 & 0.028 & \\
        UVES-5 & 4689844875502755328 & FLD-121\textunderscore100781 &  6.326978 & -71.668756 & 15.94 & 13.12 & 0.028 & *\\
        UVES-6 & 4690236267283684992 & FLD-121\textunderscore100823 &  7.173361 & -71.394368 & 16.55 & 13.70 & 0.028 & \\
        UVES-7 & 4687224842392729984 & FLD-419\textunderscore1355 & 17.095436 & -72.890707 & 16.54 & 13.76 & 0.089 & \\
        MIKE-1 & 4703711675634898816 & 2M00422381-6738579 & 10.599339 & -67.649440 & 16.27 & 13.84 & 0.015 & \\ 
        MIKE-2 & 4703647251125414272 & 2M00432921-6753477 & 10.871818 & -67.896582 & 16.19 & 13.54 & 0.016 & \\
        MIKE-3 & 4690030207641834112 & 2M00301366-7136292 &  7.556891 & -71.608113 & 16.03 & 13.25 & 0.031 & \\
        MIKE-4 & 4687923440279636736 & 2M01410043-7047414 & 25.251808 & -70.794853 & 15.84 & 13.10 & 0.025 & \\
        MIKE-5 & 4638332790102640768 & 2M02091489-7352354 & 32.312049 & -73.876518 & 15.78 & 13.16 & 0.046 & \\
        \hline
        \end{tabular}
        \tablefoot{Columns are: Gaia DR3 identification number, together with the identification number adopted in Paper I for the UVES targets and from APOGEE for the MIKE targets, Gaia coordinates, $G$ and $K_s$ magnitudes, colour excesses. Stars analysed by R21 in common with our sample are marked with *.}
    \end{table*}
    
\begin{figure}
        \centering
        \includegraphics[width=0.45\textwidth]{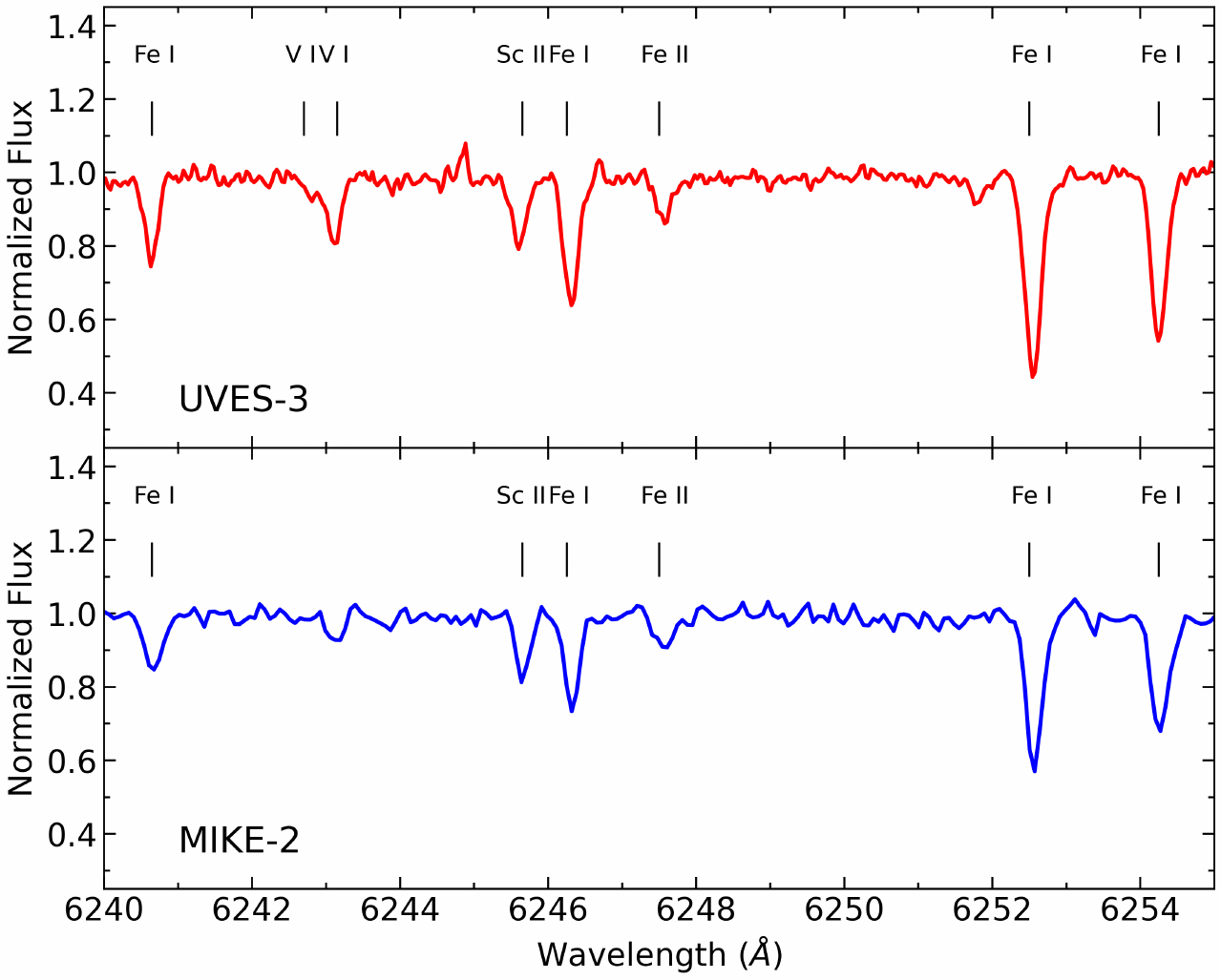}
        \caption{Comparison between UVES-3 and MIKE-2 spectra after reduction, normalisation, radial velocity and heliocentric corrections. The two stars share similar atmospheric parameters and iron abundance. Black ticks mark the position of observable lines.}
        \label{spectra}
    \end{figure}

    Other five targets were selected among the most metal-poor SMC stars from the 17th Data Release of the APOGEE survey \citep{APOGEE17} and according to the selection by \citet{Hasselquist21}. These targets were observed with the Magellan Inamori Kyocera Echelle \citep[MIKE,][]{Bernstein03, Shectman03} spectrograph on the Magellan Clay Telescope using in the 0.7" slit blue and red standard configurations (ID Programs: CN2023A-37, CN2023B-37, PI: Monaco).  This MIKE setting yields a spectral resolution of $\sim42000$ in the blue configuration and $\sim32000$ in the red one, together with a spectral coverage between 3350 \AA\;and 9500 \AA. Exposure times are 3$\times$2400s for MIKE-1 and MIKE-2, 3600s for MIKE-3, and 2700s for MIKE-4 and MIKE-5. The SNR of the combined spectra is $\sim25$ at 4500 \AA\;and $\sim40$ at 6300 \AA. 
    The spectra were reduced using the ESO pipeline\footnote{https://www.eso.org/sci/software/pipe\_aem\_table.html} for the UVES targets and the CarPy pipeline \citep{Kelson03} for the MIKE targets. All the spectra are bias-subtracted, flat-fielded, wavelength calibrated and sky-subtracted. For the UVES targets, the individual exposures were coadded together after being corrected for the appropriate heliocentric correction. Figure \ref{spectra} shows an example of UVES and MIKE  spectra used in the analysis.

\section{Spectral analysis} \label{analysis}

\subsection{Line selection} \label{linesel}
    The lines to be analysed were selected using synthetic spectra representative of each star, taking into account the appropriate stellar parameters and chemical composition, in order to identify lines with a negligible or null level of contamination from other features.\\
    The synthetic spectra were computed by means of the code SYNTHE \citep{Sbordone04, Kurucz05}. The adopted model atmospheres were calculated with ATLAS9 \citep{Castelli03} using new grids of opacity distribution functions of the KOALA database \citep{koala}. The synthetic spectra include all the atomic and molecular transitions from Kurucz/Castelli linelist, with some additional updates including the most recent laboratory measurements of $\log gf$ available in the literature (see Appendix~\ref{atomic_data}). The synthetic spectra are convolved with a Gaussian profile to reproduce the observed broadening. The latter is estimated using DAOSPEC \citep{Stetson08} and 4DAO \citep{mu13_4dao}. The instrumental profile is not able to reproduce the observed broadening that results to be larger because of macroturbulent velocity. We estimated an additional macroturbulent velocity field of $\sim$5-6 km s$^{-1}$, in line with typical values for giants stars \citep[see e.g.][]{Gray05}.\\
    We adopted the iterative scheme for the line selection described in Paper I. A first linelist was selected using synthetic spectra computed adopting the abundances of the elements measured by Paper I, a solar-scaled pattern for all the other elements apart from C and N, for which we estimated the abundances from CH and CN molecular bands, respectively (see Sect.~\ref{cn}). This initial linelist was analysed with GALA \citep{mu13} to have a first estimate of the chemical abundances of each star of the sample. Finally, a new set of synthetic spectra were computed with the entire array of abundances and they were used to select a set of lines which better resembled the chemical pattern of our targets. This procedure is especially critical for some targets with strong enhancements in neutron-capture elements, as it ensures that the specific chemical pattern is considered and prevents the inclusion of contaminated lines that would otherwise be included into the analysis if solar-scaled synthetic spectra were adopted.

\subsection{Atmospheric parameters}

    We computed the atmospheric parameters for the target stars from photometry (Table \ref{atm_par}). Effective temperatures ($T_\text{eff}$) were obtained from the broad-band colour $(G-K_s)_0$, adopting the $(G-K_s)_0-T_\text{eff}$ transformation from \cite{mu21a}, exploiting Gaia DR3 $G$ magnitudes \citep{Gaia18}, 2MASS $K_s$ magnitudes \citep{Skrutskie06} and colour excess values $E(B-V)$ from infrared dust maps \citep{Schlafly11}. The effective temperatures are characterised by uncertainties of $\sim50$ K, including the uncertainty of the calibration itself (the dominant source of error) and the errors in the $G-K_s$ colour and in the colour excess.
    
    The surface gravities ($\log g$) were derived through the Stefan-Boltzmann relation, using the $T_\text{eff}$ values, a true distance modulus $(m-M)_0=18.965\pm0.025$ \citep{Graczyk14}, bolometric corrections according to \cite{Andrae18}, and stellar masses of $0.75\pm0.10 \rm \textit{M}_\odot$. The typical $\log g$ errors are of the order of 0.1.
    
    Microturbulent velocities ($v_t$) were computed using $\log g-v_t$ relations from \cite{mu20}, consistently with Paper I. The error associated to $v_t$ is about 0.2 km s$^{-1}$, considering the quadrature sum of the uncertainties of the $\log g-v_t$ relation and of $\log g$ itself. Paper I did not determine $v_t$ values spectroscopically, in order to avoid large fluctuations, potentially caused by the small number of Fe I lines and weak lines of their GIRAFFE spectra. The new dataset of UVES and MIKE spectra allowed us to spectroscopically measure $v_t$ thanks to the large number ($\sim120-170$) of available Fe I lines. The new values are fully consistent within the uncertainties with those derived from the $\log g-v_t$ relations used above.

    \begin{table}
        \caption{RV and atmospheric parameters of the observed SMC targets.}
        \label{atm_par}
        \centering
        \begin{tabular}{c c c c c}
        \hline\hline
        ID & RV & $T_\text{eff}$ & $\log g$ & $v_t$ \\
        \hline
        & [km s$^{-1}$] & [K] & [cgs] & [km s$^{-1}$] \\
        \hline 
        UVES-1 & $137.1\pm0.5$ & 4424 & 0.95 & 1.7 \\
        UVES-2 & $156.0\pm0.5$ & 4274 & 0.59 & 2.2 \\
        UVES-3 & $150.9\pm0.5$ & 4201 & 0.49 & 2.3 \\
        UVES-4 & $119.1\pm0.5$ & 4298 & 0.72 & 1.8 \\
        UVES-5 & $106.1\pm0.5$ & 4152 & 0.43 & 1.9 \\
        UVES-6 & $179.2\pm0.5$ & 4133 & 0.66 & 1.8 \\
        UVES-7 & $139.6\pm0.5$ & 4300 & 0.71 & 1.8 \\
        MIKE-1 & $130.1\pm0.5$ & 4483 & 0.78 & 1.8 \\
        MIKE-2 & $104.1\pm0.5$ & 4282 & 0.63 & 1.9 \\
        MIKE-3 & $124.7\pm0.5$ & 4193 & 0.49 & 1.9 \\
        MIKE-4 & $226.4\pm0.5$ & 4212 & 0.43 & 1.9 \\
        MIKE-5 & $133.1\pm0.5$ & 4364 & 0.48 & 1.9 \\
        \hline
        \end{tabular}
    \end{table}

\subsection{Radial velocities} 

    We measured radial velocities (RV) using the cross correlation function (CCF) technique \citep[see e.g.][]{Tonry79} as implemented in the PyAstronomy package. Templates are synthetic spectra calculated with SYNTHE (Sect.~\ref{linesel}). The RV errors associated to this CCF measurements are of $\sim0.02$ km s$^{-1}$ and were computed by means of MonteCarlo simulations. For each target, we added Poissonian noise to the appropriate synthetic spectrum, in order to simulate the SNR of our UVES/MIKE spectra, and then Gaussian fitted the main peak of the CCF resulting from the comparison between the observed and Poissonian noise synthetic spectra, and repeated this process 200 times. The standard deviation resulting from each set of RV values was used to derive the  error on the RV measure procedure. However, the total error associated to our RV measurements is dominated by uncertainties of $\sim0.5$ km s$^{-1}$ related to the stability of UVES/MIKE and their setups. We checked the wavelength calibration accuracy of UVES/MIKE spectra by measuring the position of 5577.3 \AA\;and 6300.3 \AA\;O I sky emission lines, finding shifts compatible with a zero velocity, ruling out significant systematics in the zero-points of the wavelength scale.
    
    When compared against Paper I GIRAFFE measurements, the RV of our UVES targets show an average difference of $-1.1\pm0.5$ km s$^{-1}$. MIKE targets RV show a much better agreement with their APOGEE DR17 \citep{APOGEE17} RV measurements, with an average difference of $0.0\pm0.5$ km s$^{-1}$. Looking at single targets, four of them from both subsamples display absolute RV discrepancies larger than 1.5 km s$^{-1}$ with respect to currently available RV estimates: atmospheric jitter caused by the extended convective structure and other instabilities in the observed RGB stars can account for such discrepancies.

\subsection{Chemical abundances}

    The computation of chemical abundances for all absorption lines was done using the SALVADOR code (Alvarez Garay et al., in prep.), which performs a $\chi^2$-minimization between an observed spectrum and a grid of synthetic spectra computed on the fly with the SYNTHE code around each absorption line.  For the computation of C and N abundances for the targets observed with UVES, wider regions than the ones for single absorption lines from Blue Arm setting spectra were used, in order to correctly analyse the molecular absorption bands of CH and CN. Chemical abundances where computed using solar reference abundances from \cite{Grevesse98}, except for oxygen, for which we adopted the value by \cite{Caffau11}.  NLTE corrections are often required for chemical abundances of some elements, especially in the case of metal-poor stars \citep{Asplund05}. In our analysis, we applied NLTE corrections  by \cite{Mashonkina13} for Mg I lines, by \cite{Mashonkina17} for Ca I lines, by \cite{Mashonkina11} for Fe I lines, by \cite{Mashonkina19} for Ba II lines, by \cite{Mashonkina00} for Eu II line. All corrections were applied by means of the database from \cite{Mashonkina23}.

    Uncertainties in the computation of chemical abundances are caused by measurement errors and uncertainties from atmospheric parameters. The average abundance value from multiple absorption lines of the same element in a given star comes with a measurement error, i.e. the standard deviation of the mean abundance. These errors arise from the spectral fitting procedure and from the $\log gf$ values assumed for each absorption feature. In the case only one absorption feature of an element is available, its abundance measurement error was measured by running a Monte Carlo simulation of 200 synthetic spectra with the addition of Poissonian noise to reproduce the observed SNR. The single feature was analysed in each of these spectra with SALVADOR, and the standard deviation of the derived abundance distribution is assumed as the measurement error.
    
    The uncertainties arising from atmospheric parameters were computed with additional analyses in which atmospheric parameters were changed, according to their $1\sigma$ errors (50 K for $T_\text{eff}$, 0.1 for $\log g$, 0.2 km s$^{-1}$ for $v_t$), one at a time, in order to evaluate the effect of each parameter uncertainty on the abundance computation.

    These two sources of uncertainties were summed in quadrature to obtain the actual uncertainties associated to the measured abundances: this was done because we computed the covariance terms arising from $T_\text{eff}-\log g$ (and consequently $T_\text{eff}-v_t$) and $\log g-v_t$ relations, and we observed they impact for $<0.01$ dex on the final error of our abundances. Uncertainties in the measured iron abundances are therefore given by the following quadrature sum:
    \begin{equation}
      \sigma_{[\text{Fe/H}]}=
      \sqrt{\frac{\sigma_\text{Fe}^2}{N_\text{Fe}}+(\delta_\text{Fe}^{T_\text{eff}})^2+(\delta_\text{Fe}^{\log g})^2+(\delta_\text{Fe}^{v_t})^2}
    \end{equation}
    For all the other elements, since their abundances are in the form [X/Fe], their uncertainties were consequently computed in the following way:
    \begin{multline}
      \sigma_{[\text{X/Fe}]}=\\
      \sqrt{\frac{\sigma_\text{X}^2}{N_\text{X}}+\frac{\sigma_\text{Fe}^2}{N_\text{Fe}}+(\delta_\text{X}^{T_\text{eff}}-\delta_\text{Fe}^{T_\text{eff}})^2+(\delta_\text{X}^{\log g}-\delta_\text{Fe}^{\log g})^2+(\delta_\text{X}^{v_t}-\delta_\text{Fe}^{v_t})^2}
    \end{multline}
    In the two above formulas $\sigma_\text{Fe}$ and $\sigma_\text{X}$ are the standard deviations of Fe and X element abundances, respectively; $N_\text{X}$ and $N_\text{Fe}$ are the number of fitted absorption lines used in the abundance measurement of Fe and the X element, respectively; $\delta_\text{X, Fe}^{p}$ are the abundance variations caused by the variation of the atmospheric parameter $p$.

\section{Iron abundances}\label{iron}

    The iron abundances of the 12 SMC stars were derived from $\sim150-170$ Fe~I lines. We measured also 15-20 Fe~II lines, but in two metal-poor MIKE targets only 5 Fe~II lines are available. Table~\ref{Fe} lists the average LTE and NLTE Fe abundances derived from neutral lines, and the LTE abundance from single ionised lines (for the latter the NLTE effects are negligible because Fe~II is the dominant ionization stage). Because  of the excellent agreement between the abundances from Fe~I and Fe~II lines and the higher precision of the [Fe/H] derived from the neutral lines (due to the huge number of used lines), throughout the paper we refer only to the NLTE [Fe/H] from Fe~I lines.

    \begin{table}[h!]
        \caption{Fe I LTE, Fe I NLTE and Fe II.}
        \label{Fe}
        \centering
        \begin{tabular}{c c c c}
        \hline\hline
        Target & [FeI/H] LTE & [FeI/H] NLTE & [FeII/H] \\
        \hline 
        UVES-1 & $-1.92\pm0.08$ & $-1.84\pm0.08$ & $-1.80\pm0.11$ \\
        UVES-2 & $-2.30\pm0.09$ & $-2.20\pm0.09$ & $-2.17\pm0.12$ \\
        UVES-3 & $-2.20\pm0.08$ & $-2.12\pm0.08$ & $-2.10\pm0.11$ \\
        UVES-4 & $-1.95\pm0.09$ & $-1.87\pm0.09$ & $-1.86\pm0.12$ \\
        UVES-5 & $-2.14\pm0.08$ & $-2.08\pm0.08$ & $-2.00\pm0.12$ \\
        UVES-6 & $-1.40\pm0.05$ & $-1.36\pm0.05$ & $-1.57\pm0.12$ \\
        UVES-7 & $-2.06\pm0.09$ & $-1.97\pm0.09$ & $-1.94\pm0.12$ \\
        MIKE-1* & $-2.34\pm0.11$ & $-2.21\pm0.11$ & $-1.83\pm0.22$ \\
        MIKE-2 & $-2.27\pm0.09$ & $-2.17\pm0.09$ & $-2.19\pm0.13$ \\
        MIKE-3* & $-2.30\pm0.12$ & $-2.22\pm0.12$ & $-2.01\pm0.17$ \\
        MIKE-4 & $-1.90\pm0.09$ & $-1.83\pm0.09$ & $-1.85\pm0.13$ \\
        MIKE-5 & $-1.60\pm0.09$ & $-1.50\pm0.09$ & $-1.70\pm0.13$ \\
        \hline
        \end{tabular}
        \tablefoot{Targets with * have Fe II based only on 5 Fe II lines, the others are generally based on 15-20 Fe II lines.}
    \end{table}

    \begin{figure}
        \centering
        \includegraphics[width=0.45\textwidth]{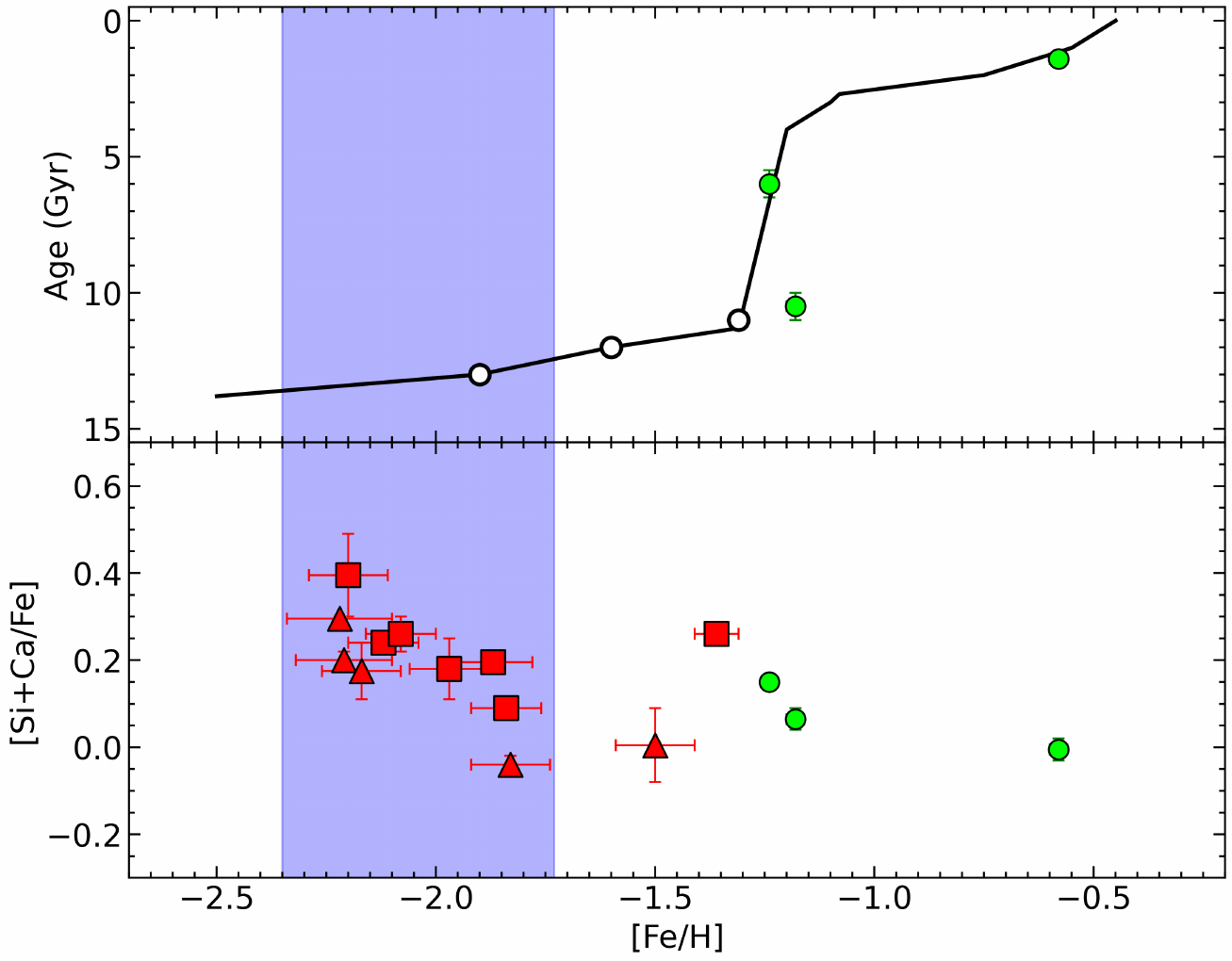}
        \caption{\textit{Upper panel}: theoretical AMR of the SMC \citep[black line,][]{Pagel98}, superimposed with the SMC GCs (Paper II, green circles). The white dots mark the position on the theoretical AMR of ages 11, 12 and 13 Gyr. \textit {Lower panel}: average Si and Ca abundances as a function of [Fe/H] for SMC stars of this work observed with UVES (red squares) and MIKE (red triangles). The blue shaded area highlights the [Fe/H] region of the stars of our sample with $-2.2<$[Fe/H]$<-1.8$ dex.}
        \label{AMR}
    \end{figure}
    
    Our sample of SMC stars has $-2.3<$[Fe/H]$<-1.4$ dex, with 10 out of 12 stars with [Fe/H]$<-1.8$ dex. The age-metallicity relation (AMR) of \cite{Pagel98} allows us to provide an estimate of the age of these stars (see Fig.~\ref{AMR}). In fact, according to this AMR, all our targets should be older than $\sim11$ Gyr. In particular, the two metal-rich targets should have an age between 11 and 12 Gyr, while the other 10 stars should be formed in the first 1.5 Gyr of life of the galaxy. As a reference, we recall that the oldest SMC GC, NGC~121, has an age of 10.5$\pm$0.5 Gyr \citep{Glatt08}, with [Fe/H]$=-1.18\pm0.02$ dex (Paper II). Moreover, since all our targets show [Fe/H]$<-1.4$ dex, their formation should have happened before the SMC quiescent phase, during which a wide spread in ages, from 11 to 4 Gyr, corresponds to a rather small variation of iron, $-1.3<$[Fe/H]$<-1.2$ dex, as a consequence of a low SFR, while this galaxy was also characterised by an inhomogeneous evolution.
    
    The four SMC metal-poor stars analysed by R21 are in common with our sample (see Table \ref{id_coord}). We compare our atmospheric parameters and [Fe/H] with those by R21. In their work, R21 determined both $T_\text{eff}$ and $v_t$ of their targets spectroscopically, while $\log g$ were derived using isochrones. The  average differences are ($<T_\text{eff}-T_\text{eff, R21}>=-130\pm60$ K), ($<\log g-\log g_\text{R21}>=-0.10\pm0.06$ in cgs units) and ($<v_t-v_{t,\text{ R21}}>=-0.99\pm0.18$ km s$^{-1}$). The latter difference is particularly significative and we attribute their large values of $v_t$ ($3.0$ km s$^{-1}$ on average, unlikely for low-mass giant stars) to the spectroscopical determination of this parameter that is based on the balance of the abundances from weak and strong lines. The MIKE spectra analysed by R21 have relatively low SNR ($\sim45$ at 6500 \AA, $\sim20$ at 4500 \AA, on average), thus introducing a bias against the weak lines that leads to spurious $v_t$.
    
    There is a significant offset of $<$[Fe/H]$-$[Fe/H]$_\text{R21}>=+0.25\pm0.06$ dex between the two sets of abundances. Given the magnitude of NLTE corrections on Fe I lines, R21 preferred to use abundances from Fe II lines as their final [Fe/H] measurements. Even considering their [Fe/H] from Fe I lines, we still get a significant offset of $<$[Fe/H]$-$[Fe/H]$_\text{R21}>=+0.26\pm0.08$ dex, though in this case we are comparing our set of NLTE Fe I abundances with a LTE one. If we compare the two sets of LTE Fe I abundances, then the offset actually reduces to $<$[Fe/H]$_\text{LTE}-$[Fe/H]$_\text{R21}>=+0.18\pm0.08$ dex. We attribute this difference to the differences in $v_t$ and $T_\text{eff}$.
    
    The iron abundances we measured for UVES targets are compatible with their previous GIRAFFE measurements (Paper I), with an average difference $<$[Fe/H]$_\text{UVES}-$[Fe/H]$_\text{GIRAFFE}>=0.10\pm0.08$ dex. UVES-6 is the only target showing a significant difference ([Fe/H]$_\text{UVES}-$[Fe/H]$_\text{GIRAFFE}=0.55\pm0.15$ dex), likely due to the low spectral quality of the GIRAFFE spectra for this star. Finally, we compare [Fe/H] for MIKE targets with the values available in APOGEE DR17, finding an excellent agreement, with an average difference $<$[Fe/H]$_\text{MIKE}-$[Fe/H]$_\text{APOGEE}>=+0.04\pm0.07$ dex.

\section{Chemical abundance ratios}\label{abundances}

    Here we present chemical abundances for C, N, $\alpha$-elements (O, Mg, Si, Ca) and neutron-capture elements (Y, Ba, La, Ce, Nd, Sm, Eu). For the latter, Ba and La were already measured in Paper I for the UVES targets, while no abundances are available from APOGEE for the MIKE targets. We show the abundances of the targets in a series of [X/Fe]-[Fe/H] abundance diagrams (Figs.~\ref{alpha}-\ref{Nd}), in comparison with the abundances of the SMC field stars (Paper I) and of the SMC globular clusters (Paper II). Additionally, we show abundance ratios for MW field stars from SAGA database \cite{Suda08}. Since this database compiles results from different analyses, it is not fully homogeneous and should be considered in a statistical sense. Nevertheless, the overall trends of MW stars are a useful reference for a comparison with the SMC stars, despite this comparison can be affected by the systematics among the different analyses (in terms of model atmospheres, solar abundance values, NLTE corrections, line lists, and use of dwarf and giant stars).

\subsection{Carbon and Nitrogen}\label{cn}

    For cool stars as the targets analysed here, the spectrum can be contaminated by several CN or CH molecular features, therefore the use of appropriate C and N abundances is necessary in the evaluation of the level of blending of each feature. For the UVES targets only we measured C abundances from the CH G-band at 4300 \AA\;and N abundances from the CN molecular band at 3880 \AA. For the MIKE targets, the SNR at the wavelengths of the CH and CN molecular bands is too low to allow us a reliable measure, hence we adopted [C/Fe] and [N/Fe] average values measured in the UVES targets for the spectral synthesis of the other elements. Since the main focus of this analysis is on $\alpha$ and neutron capture elements, we provide more details about C and N abundances of our targets in Appendix~\ref{appendix_b}.
 
\subsection{\texorpdfstring{$\alpha$}{}-elements}
\label{alfa}

    \begin{figure*}
        \centering
        \includegraphics[width=0.45\textwidth]{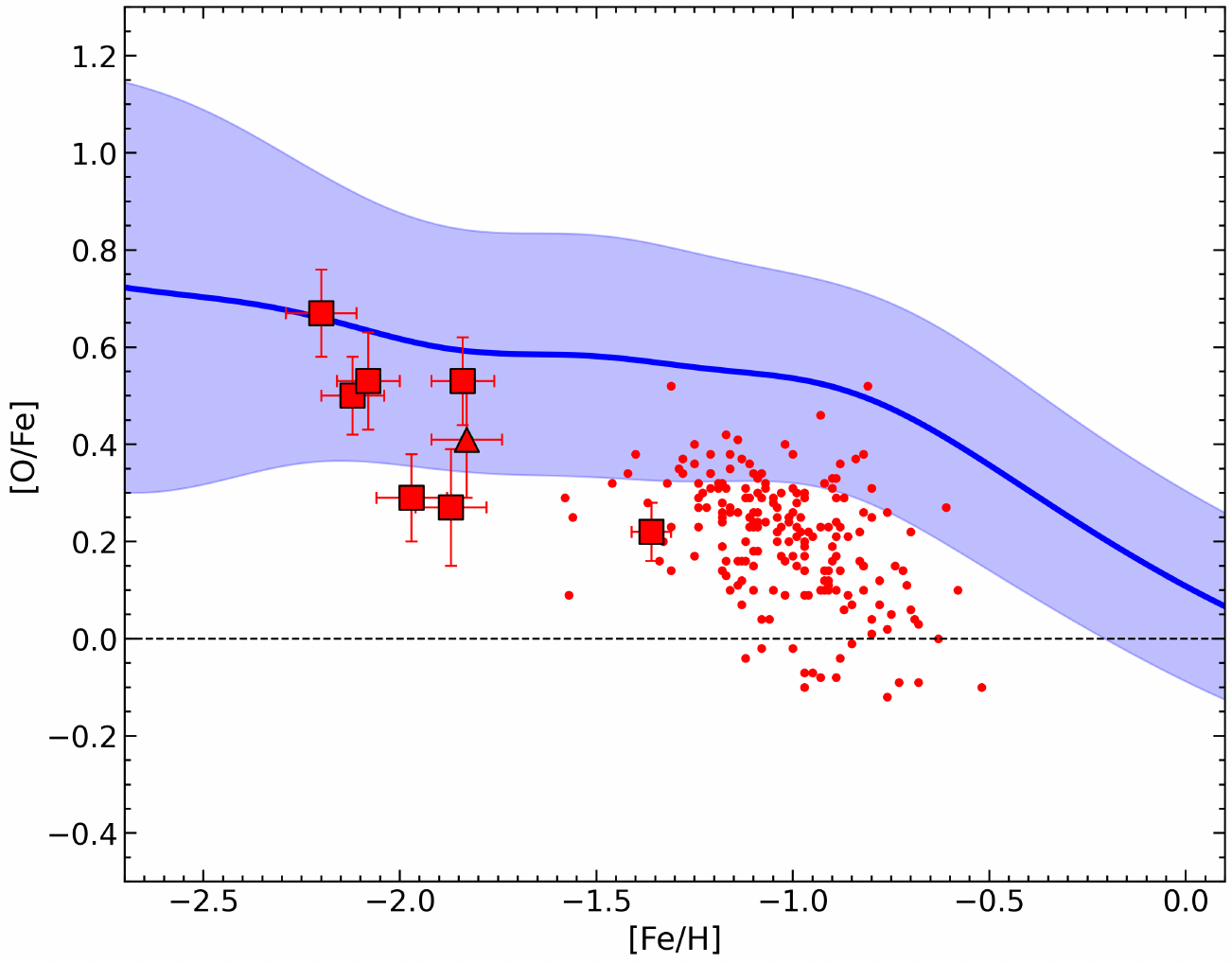}
        \includegraphics[width=0.45\textwidth]{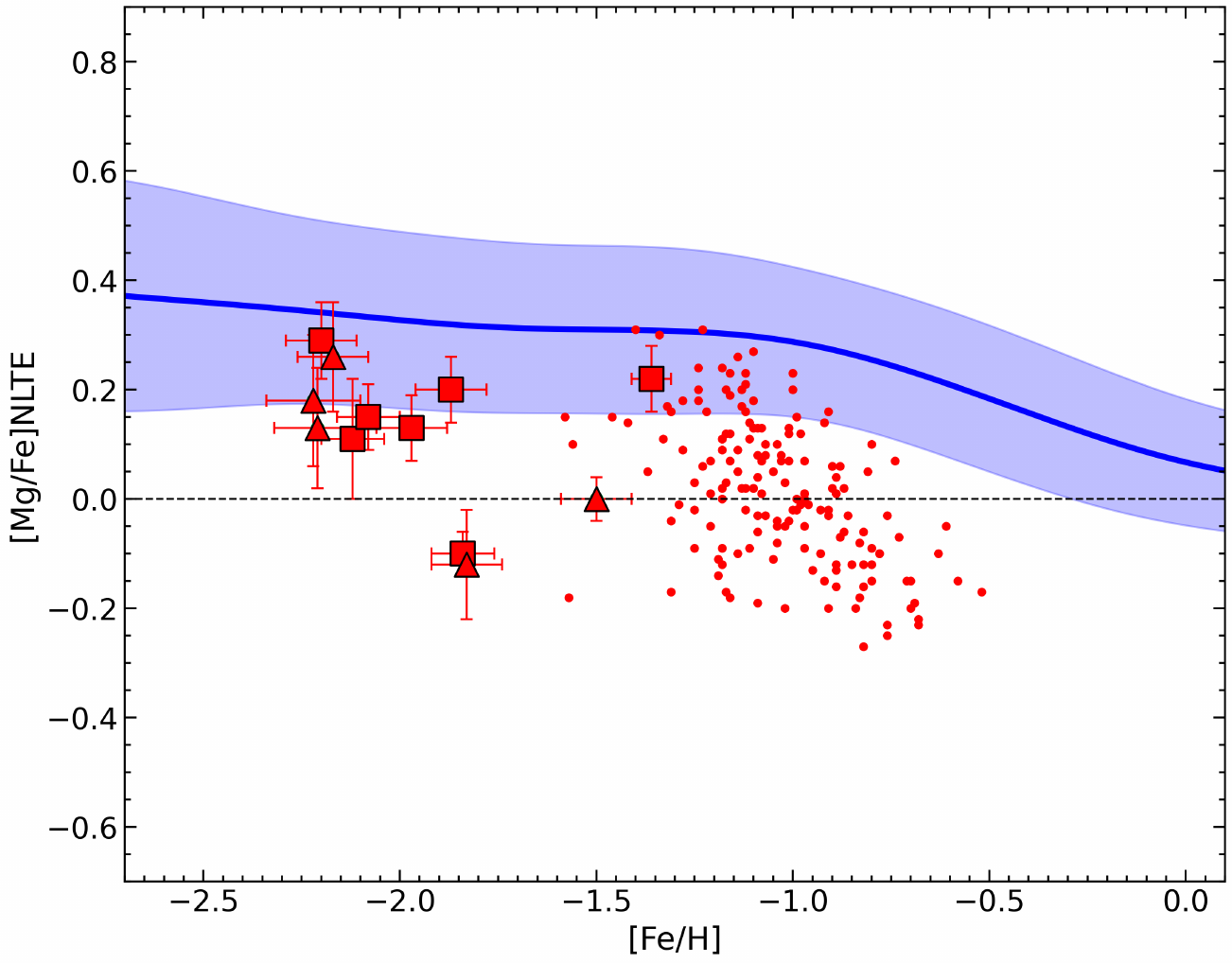}
        \includegraphics[width=0.45\textwidth]{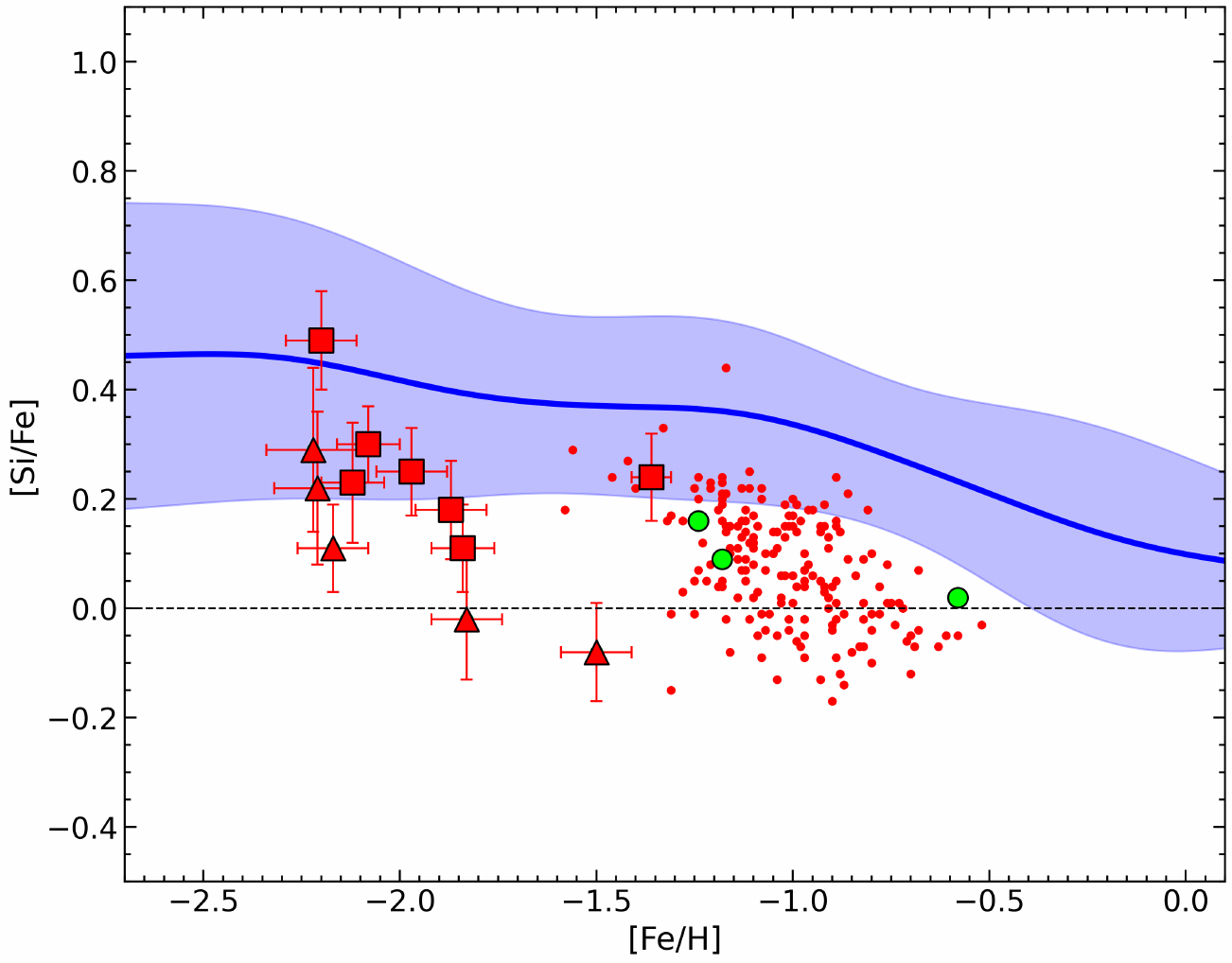}
        \includegraphics[width=0.45\textwidth]{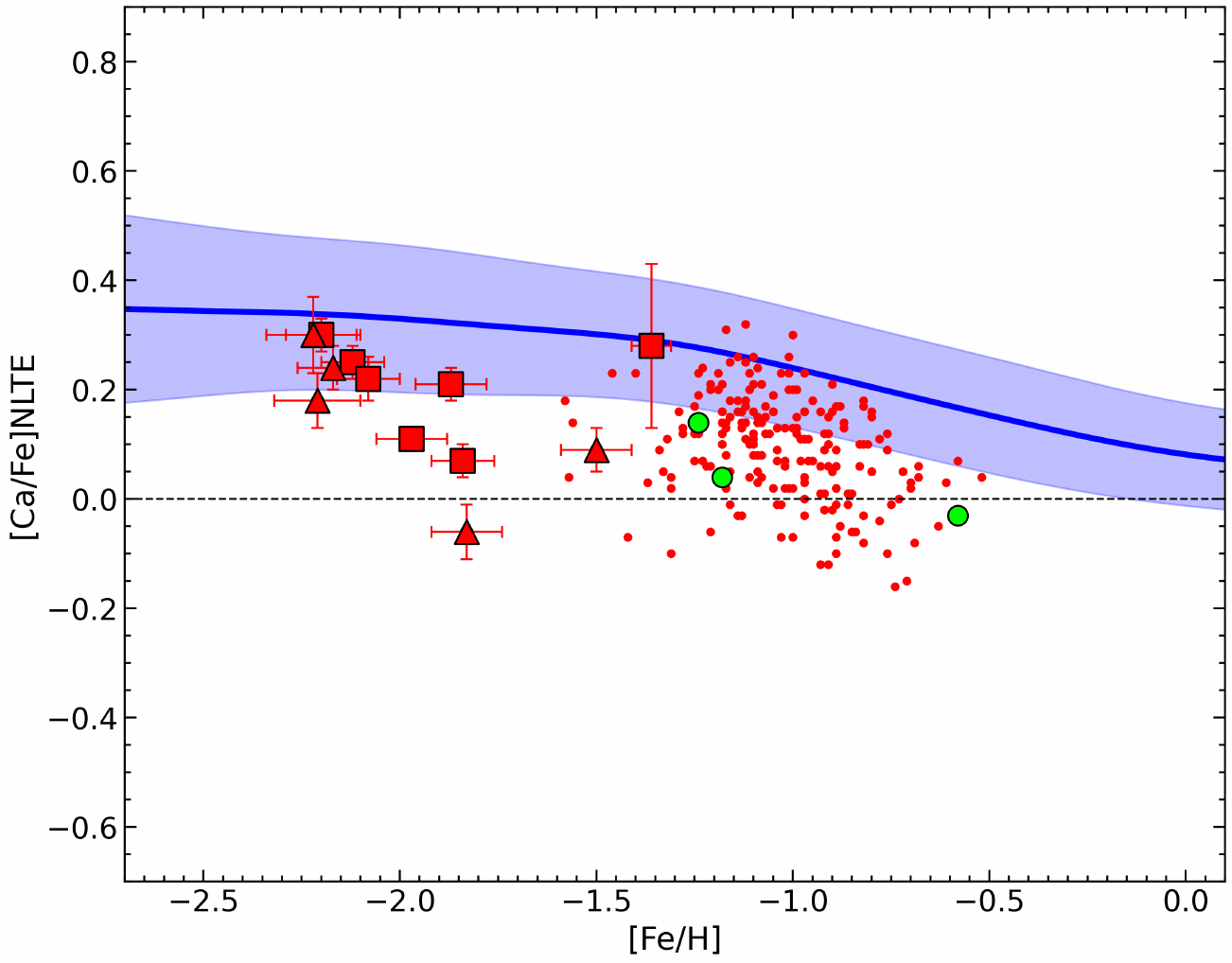}
        \caption{$\alpha$-elements patterns: [O/Fe], [Mg/Fe], [Si/Fe], [Ca/Fe] abundances as a function of [Fe/H] for SMC stars of this work observed with UVES (red squares) and MIKE (red triangles), together with the SMC field stars (Paper I, red small circles) and the SMC GCs (Paper II, green circles). Error bars are shown for UVES and MIKE abundances. As a reference, Gaussian KDE regressions of MW abundances built from SAGA database \citep{Suda08} (blue line) are reported, together with their 1$\sigma$ confidence interval (blue shaded area), in the background.}
        \label{alpha}
    \end{figure*} 
    We measured abundances for the $\alpha$-elements O, Mg, Si and Ca. Oxygen abundances were computed from the forbidden line at 6300.3 \AA, cleaned from possible telluric contamination using appropriate synthetic spectra of the Earth atmosphere calculated with the tool TAPAS \citep{bertaux14}. Magnesium abundances were obtained from the Mg lines at 4167.3 \AA\;and 5711.1 \AA. Silicon abundances were based on the absorption features at 5948.5 \AA\;and 6155.1 \AA\ . Calcium abundances were derived from $\sim15$ Ca I lines. Figure~\ref{alpha} displays the run of the above mentioned [$\alpha$/Fe] abundance ratios as a function of [Fe/H]. All the targets exhibit mildly enhanced [$\alpha$/Fe] abundance ratios, about $\sim+0.2$ dex, apart from [O/Fe] that generally show higher values. In general, the SMC stars at these [Fe/H] have lower [$\alpha$/Fe] than those measured in MW stars of similar [Fe/H], as found both from optical (Paper I) and near-infrared analyses \citep{Nidever20,Hasselquist21,Hasselquist24}.

\subsection{\texorpdfstring{$r$}{}-process elements}

    \begin{figure}
        \centering
        \includegraphics[width=0.45\textwidth]{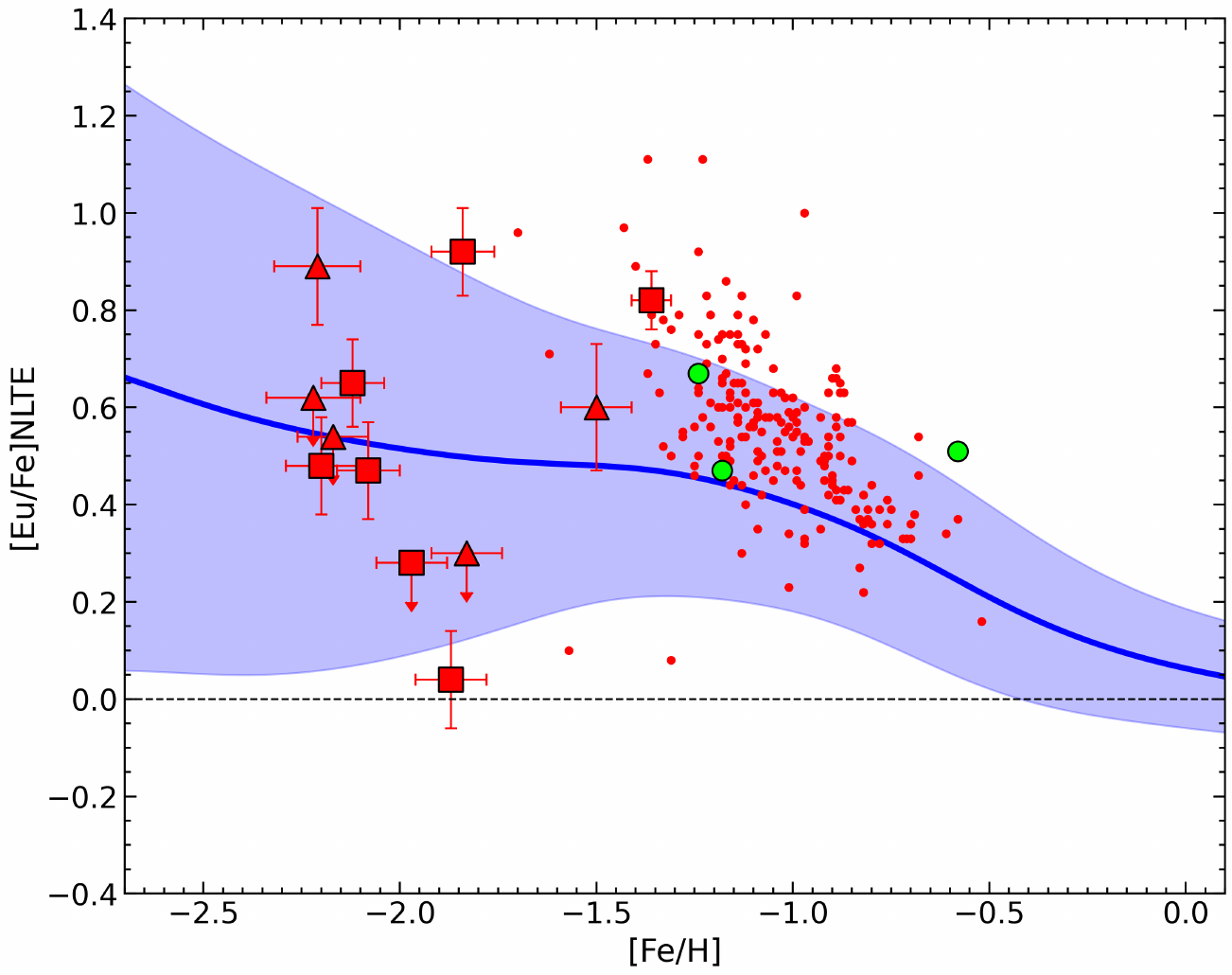}
        \caption{[Eu/Fe] abundances as a function of [Fe/H]. Same symbols of Fig.~\ref{alpha}, but the red small circles are SMC field stars observed with GIRAFFE from Paper IV.}
        \label{Eu}
    \end{figure}

    \begin{figure}
        \centering
        \includegraphics[width=0.45\textwidth]{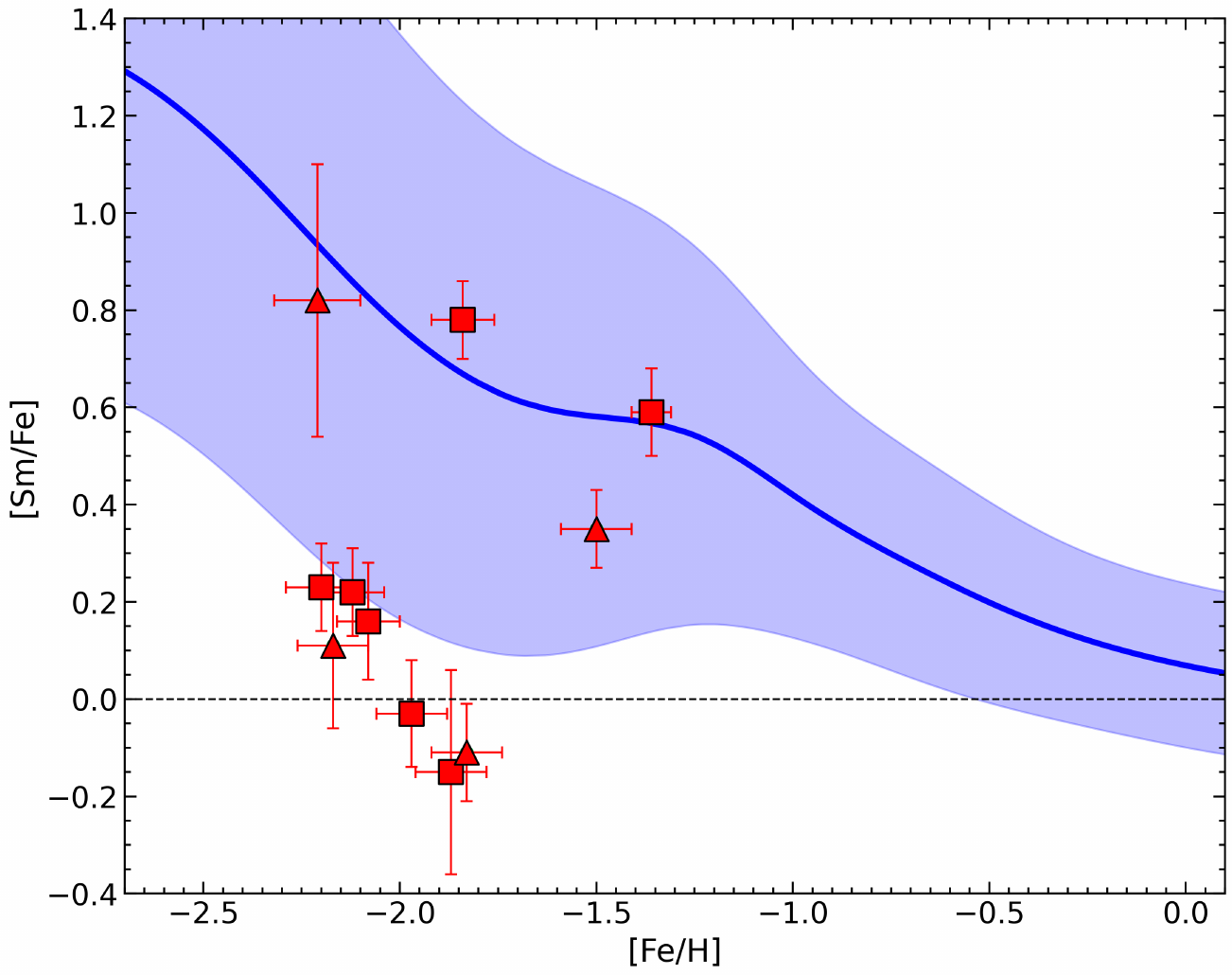}
        \caption{[Sm/Fe] abundances as a function of [Fe/H]. Same symbols of Fig.~\ref{alpha}.}
        \label{Sm}
    \end{figure}

    We measured two neutron capture elements mainly formed by $r$-process, namely europium (one of the purest $r$-process elements) and samarium, produced by $r$-process at $\simeq65$ \% in the solar system \citep{Sneden08}.
    
    Europium: We computed Eu abundances from the Eu II line at 6645.1 \AA\ (and from 6437.6 \AA\ when visible). As shown in Fig.~\ref{Eu}, our [Eu/Fe] abundances are indeed comparable with MW stars Eu content. The SMC targets also exhibit a large star-to-star scatter in [Eu/Fe], spanning $\sim1$ dex and larger than their uncertainties, which indicates an intrinsic scatter among our stars, in line with the first claim provided in Paper IV.
    
    All [Eu/Fe] abundance ratios are supersolar, with the three highest [Eu/Fe] stars (UVES-1, UVES-6, MIKE-1) that can be classified as $r$-II stars (namely, with [Eu/Fe]$>+0.7$, \citealt{Holmbeck20}). These three $r$-II stars confirm that $r$-process synthesis can be extremely efficient in a galaxy like the SMC, as already suggested by R21. Finally, we compare our [Eu/Fe] with those by R21 for the stars UVES-1 and UVES-5, finding an average difference $<$[Eu/Fe]$-$[Eu/Fe]$_\text{R21}>=-0.23\pm0.15$ dex.

    Samarium: We identified few Sm lines, always blueward of 4950 \AA. The [Sm/Fe] abundances we measure (Fig.~\ref{Sm}) display a $\sim1$ dex wide star-to-star scatter as observed for [Eu/Fe]. In particular, 4 stars have high [Sm/Fe] (the same stars are also among the most Eu-rich targets), while the other stars shows a mild decrease of [Sm/Fe] by increasing [Fe/H]. As expected due to the common $r$-process origin, we find correspondence between Sm and Eu contents: the stars with the highest (lowest) [Sm/Fe] abundances are those with the highest (lowest) [Eu/Fe].
    
\subsection{\texorpdfstring{$s$}{}-process elements} \label{s}

    We measured some neutron-capture elements mainly formed through $s$-process at solar metallicity, namely Y (belonging to the first peak of $s$-process), and Ba, La, Ce and Nd (belonging to the second peak).

\begin{figure}
        \centering
        \includegraphics[width=0.45\textwidth]{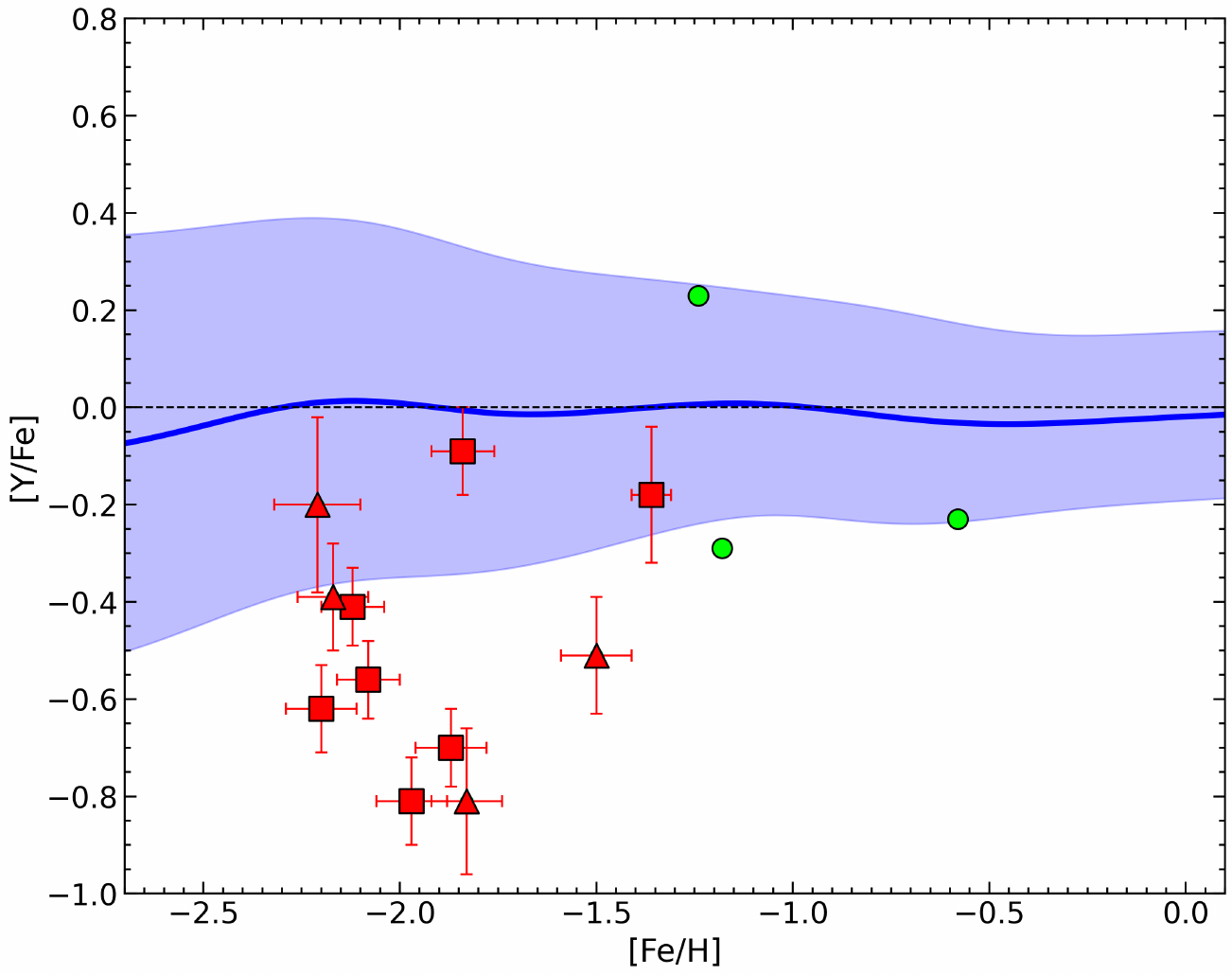}
        \caption{[Y/Fe] abundances as a function of [Fe/H]. Same symbols of Fig.~\ref{alpha}.}
        \label{Y}
    \end{figure}

    Yttrium: The yttrium abundances are based on 4 to 12 Y II lines, depending on the target. Figure~\ref{Y} shows the SMC stars [Y/Fe] abundances that appear to be, on average, lower than the MW stars of similar [Fe/H].  
    
\begin{figure}
        \centering
        \includegraphics[width=0.45\textwidth]{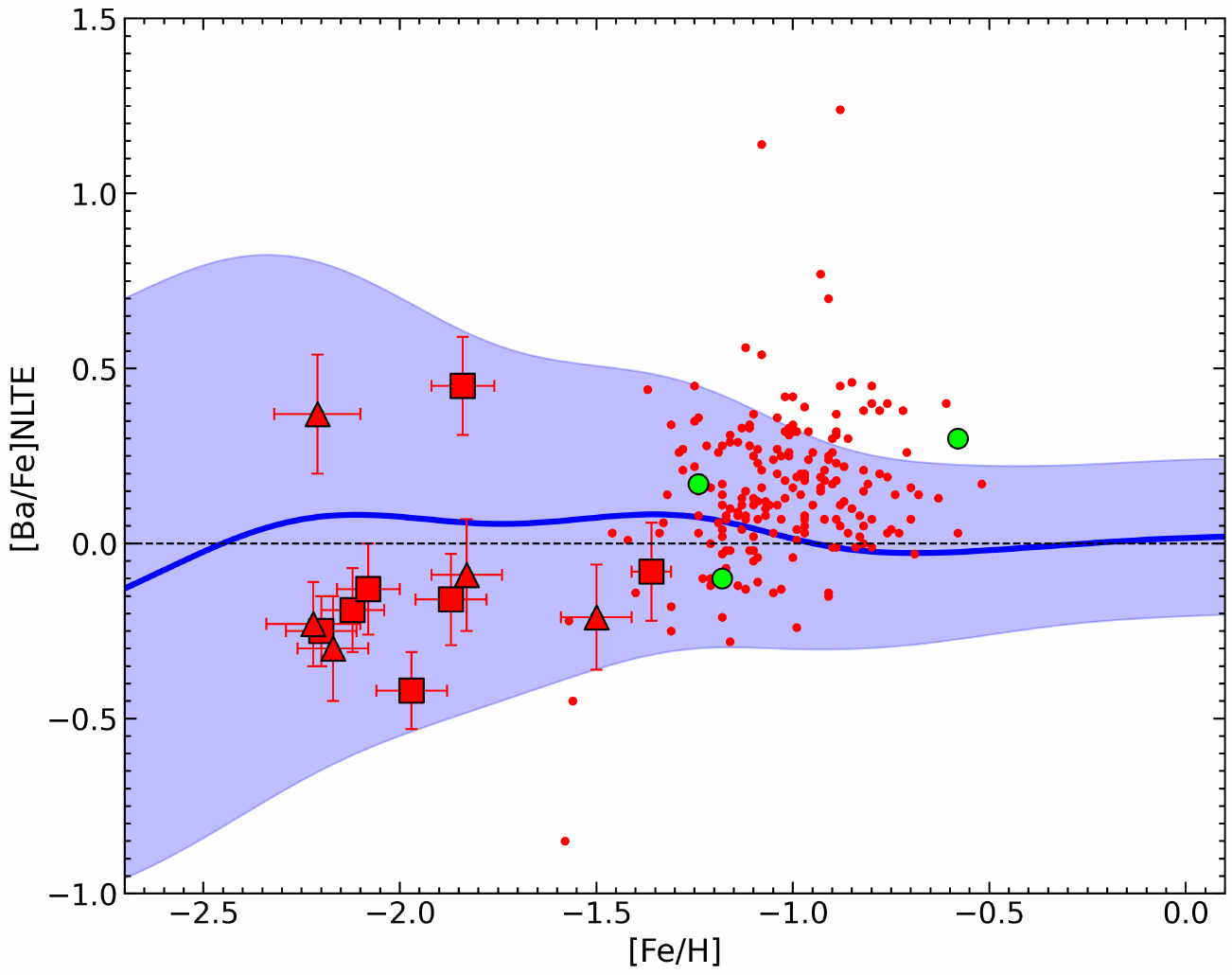}
        \caption{[Ba/Fe] abundances as a function of [Fe/H]. Same symbols of Fig.~\ref{alpha}.}
        \label{Ba}
    \end{figure}

    Barium: barium is a second-peak $s$-process element, mainly produced in AGB stars generally below $4\;{\rm \textit{M}_\odot}$ \citep{Gallino98}, with yields highly dependent on metallicities, and only in  minor amount in fast rotating massive stars. We measured [Ba/Fe] from Ba II lines at 4934.1 \AA, 5853.7 \AA, 6141.7 \AA, 6496.9 \AA. Since some of these features are often saturated, the abundances obtained from their profiles should be handled with care and their uncertainties are more sensitive to the adopted values, in particular of microturbulence, in model atmosphere. Figure~\ref{Ba} highlights that our [Ba/Fe] are indeed comparable with Galactic values. However, two stars (UVES-1 and MIKE-1) stand out from other SMC metal-poor stars, at clearly higher [Ba/Fe]: we point out how UVES-1 and MIKE-1 also are the two stars with the highest [Eu/Fe].

\begin{figure}
        \centering
        \includegraphics[width=0.45\textwidth]{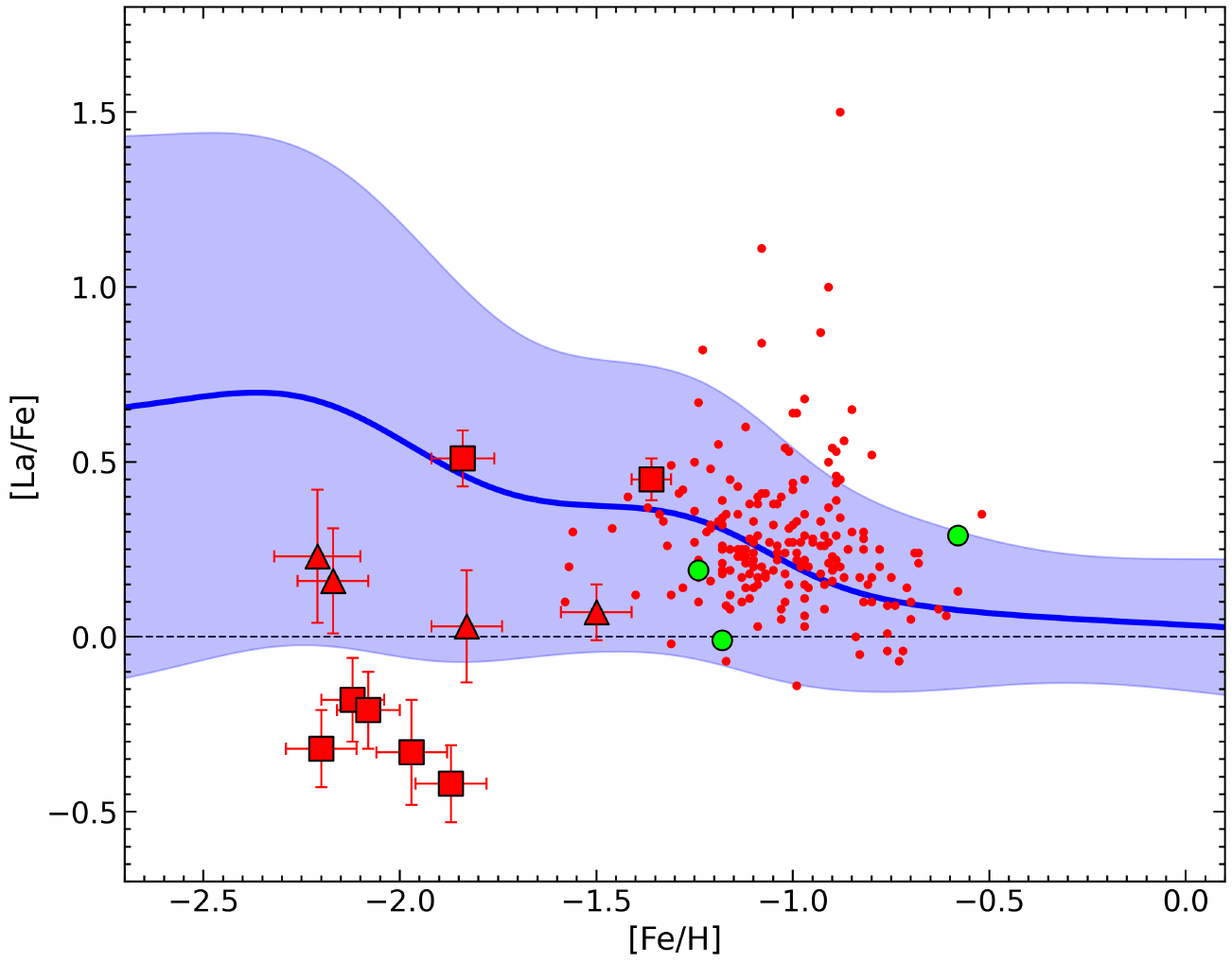}
        \caption{[La/Fe] abundances as a function of [Fe/H]. Same symbols of Fig.~\ref{alpha}.}
        \label{La}
    \end{figure}

    Lanthanum: We measured lanthanum abundances relying on $\sim10$ La II lines in UVES spectra, and only few lines in MIKE spectra. Contrarily to barium, nearly half of the sample is depleted in [La/Fe] with respect to Galactic values (Fig.~\ref{La}). Five out of the six stars with the highest (supersolar) [La/Fe] have [Eu/Fe]$>+0.5$: the stars with the highest [La/Fe] are indeed the stars with the highest [Eu/Fe]. 

\begin{figure}
        \centering
        \includegraphics[width=0.45\textwidth]{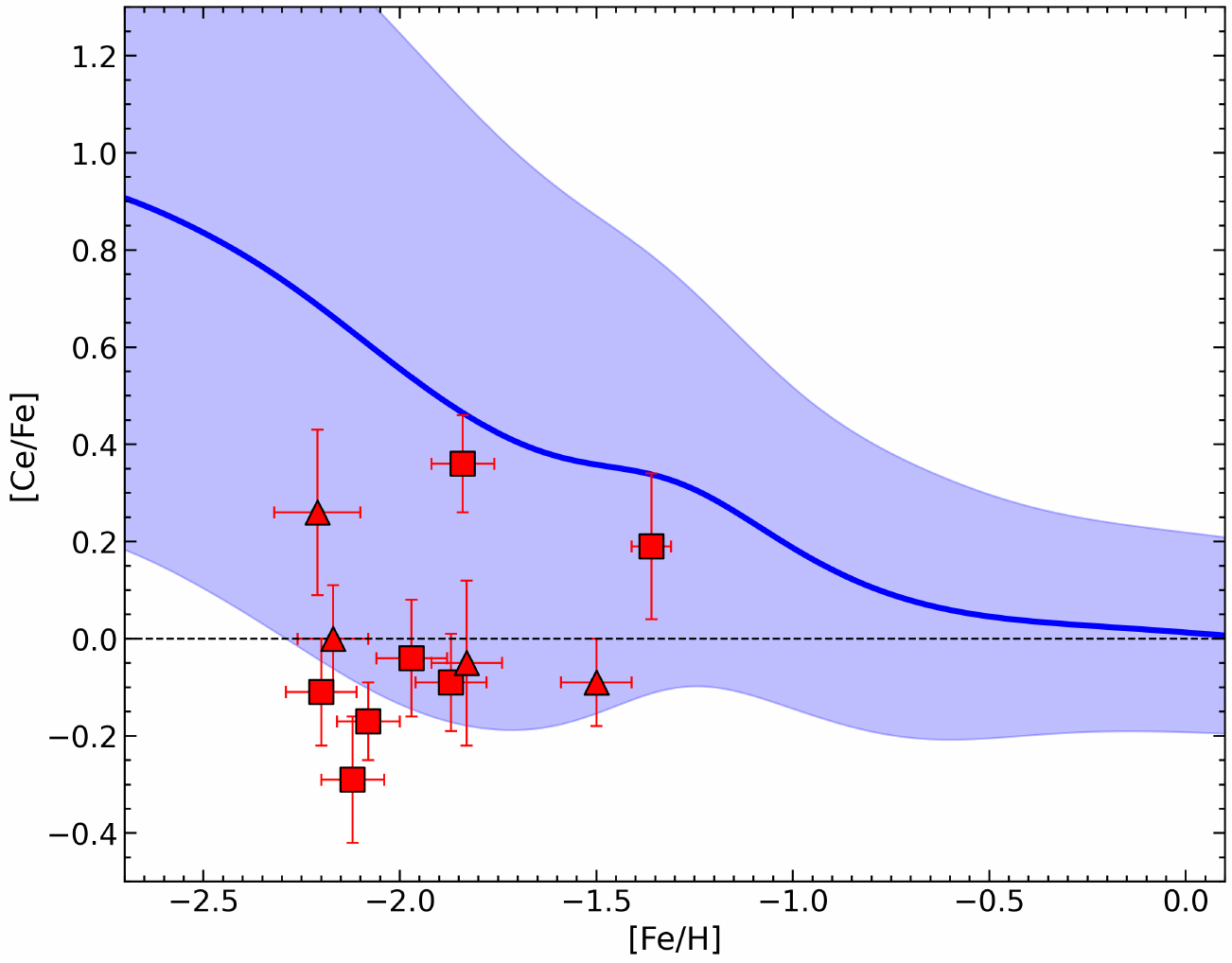}
        \caption{[Ce/Fe] abundances as a function of [Fe/H]. Same symbols of Fig.~\ref{alpha}.}
        \label{Ce}
    \end{figure}

    Cerium: Few Ce II lines are visible, mainly in regions blueward of 4650 \AA. The [Ce/Fe] abundances of 8 out of 12 metal-poor SMC stars are depleted with respect to MW stars at similar [Fe/H] (Fig.~\ref{Ce}). Most of our targets have subsolar [Ce/Fe] values, comparable with APOGEE [Ce/Fe] abundances \citep{Hasselquist21}. Once again, the three stars with supersolar and highest [Ce/Fe] of our sample (UVES-1, UVES-6, MIKE-1) are indeed the three $r$-II stars we previously mentioned.

\begin{figure}
        \centering
        \includegraphics[width=0.45\textwidth]{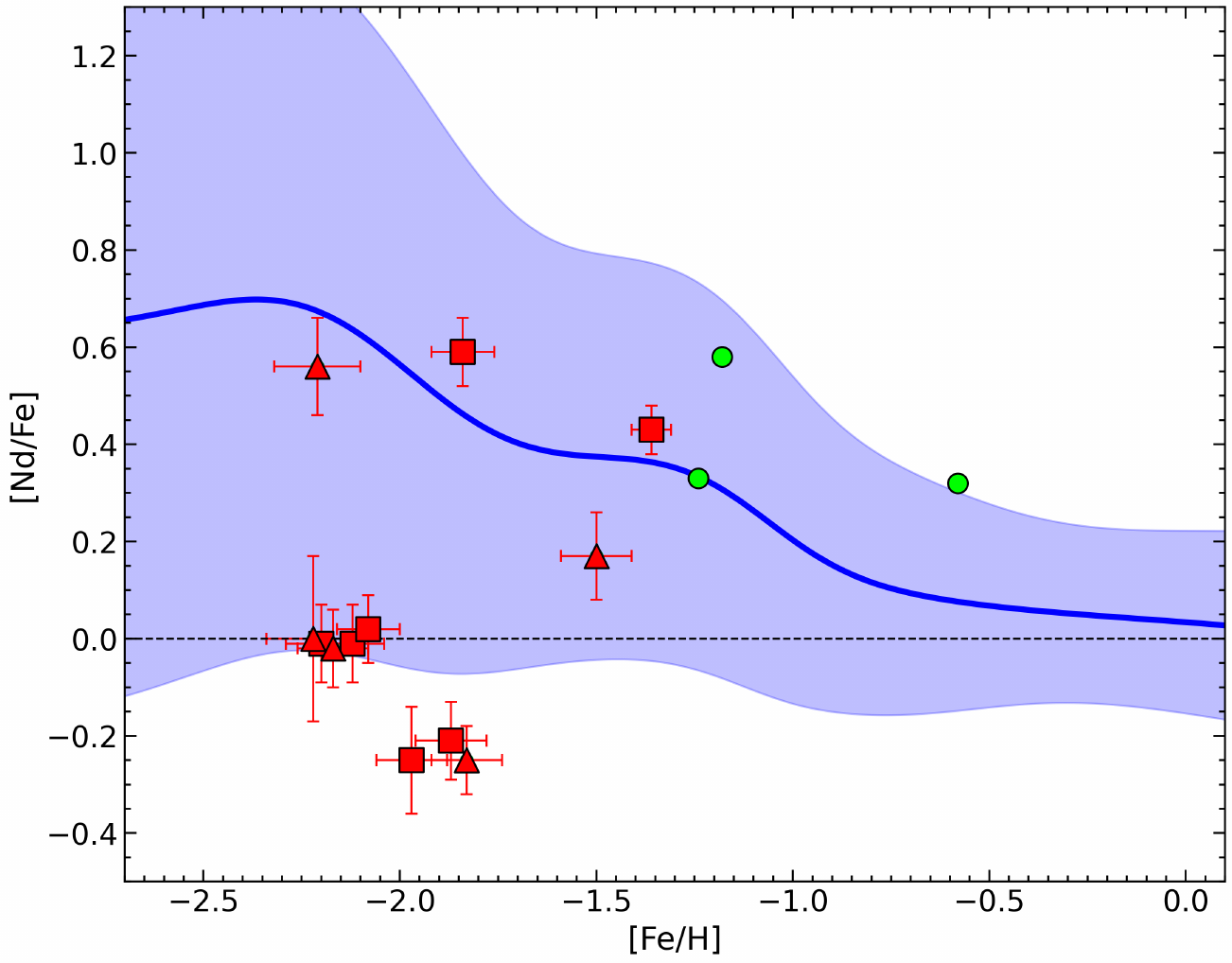}
        \caption{[Nd/Fe] abundances as a function of [Fe/H]. Same symbols of Fig.~\ref{alpha}.}
        \label{Nd}
    \end{figure}

    Neodymium: We determined neodymium abundances from $\sim15$ Nd II lines per UVES spectra and only few per MIKE spectra. The obtained [Nd/Fe] abundances are generally solar or subsolar and slightly lower than Galactic stars of similar [Fe/H] (Fig.~\ref{Nd}). Again, the four stars (UVES-1, UVES-6, MIKE-1, MIKE-5) with the highest [Nd/Fe] all have [Eu/Fe]$>$+0.6, while the three stars (UVES-4, UVES-7, MIKE-4) with the lowest [Nd/Fe] are indeed the stars with the lowest [Eu/Fe] of the entire SMC sample.

\section{The early chemical enrichment of the SMC}
\label{ss:rproc_sproc}

    \begin{figure*}
        \centering
        \includegraphics[width=0.40\textwidth]{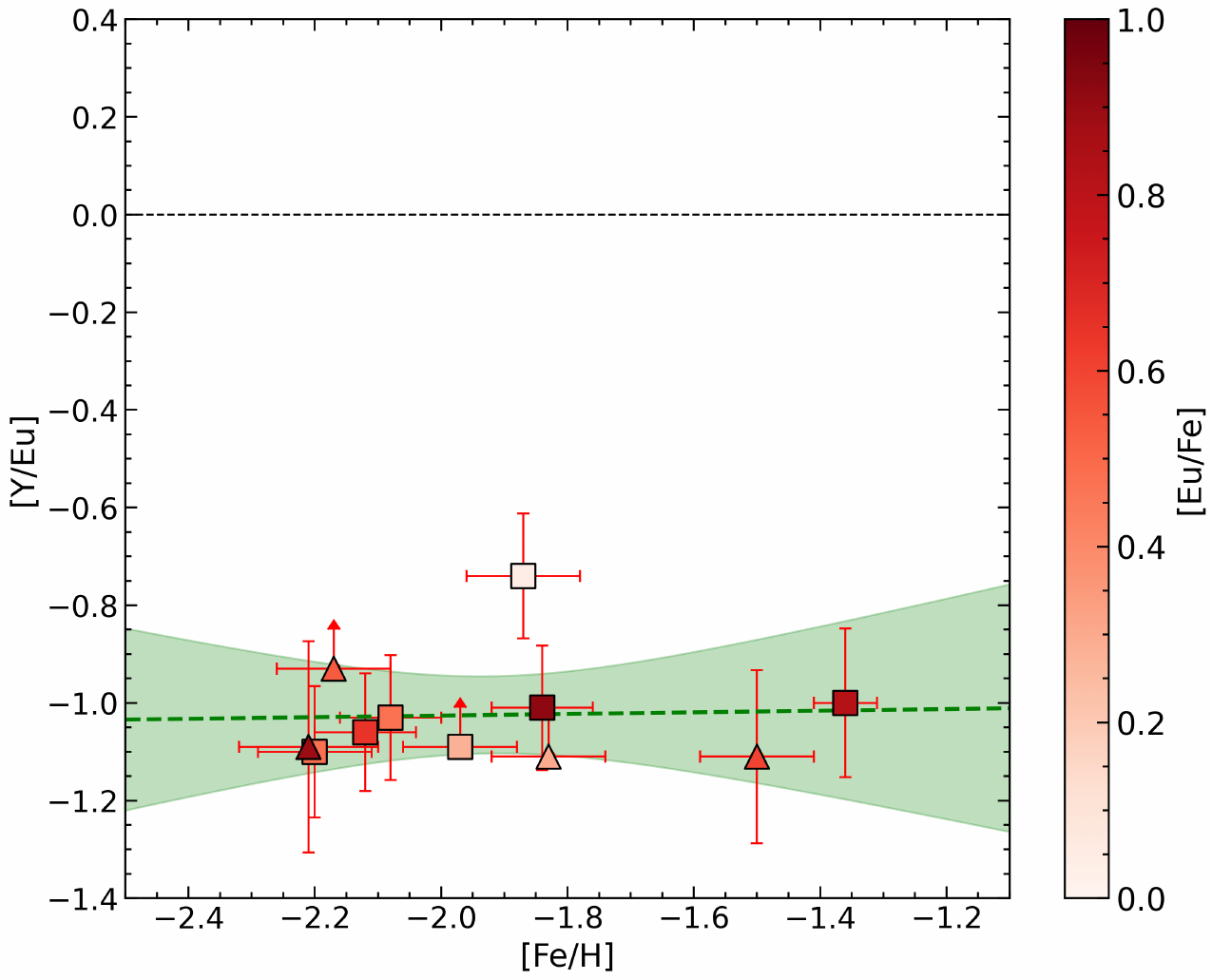}
        \includegraphics[width=0.40\textwidth]{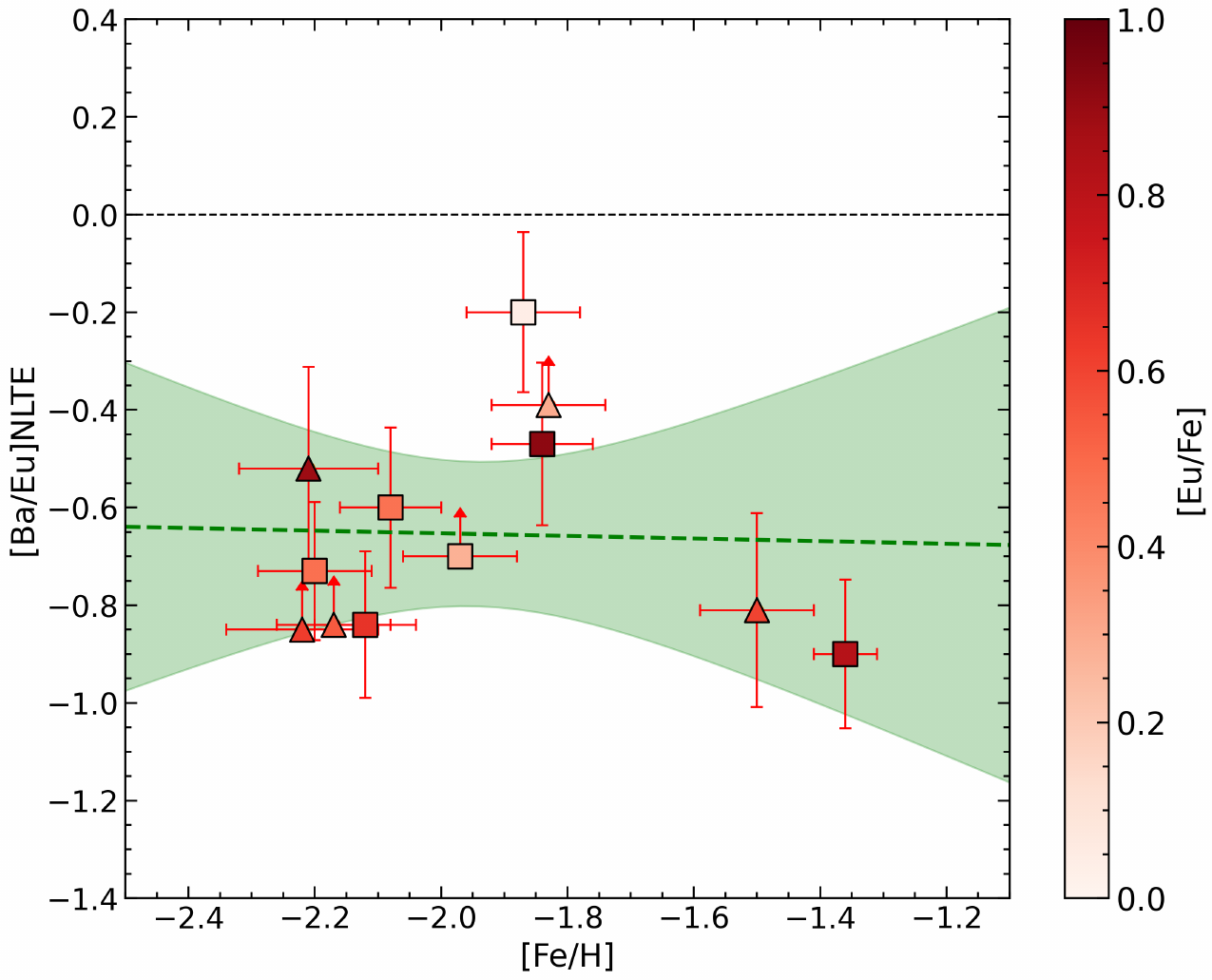}
        \includegraphics[width=0.40\textwidth]{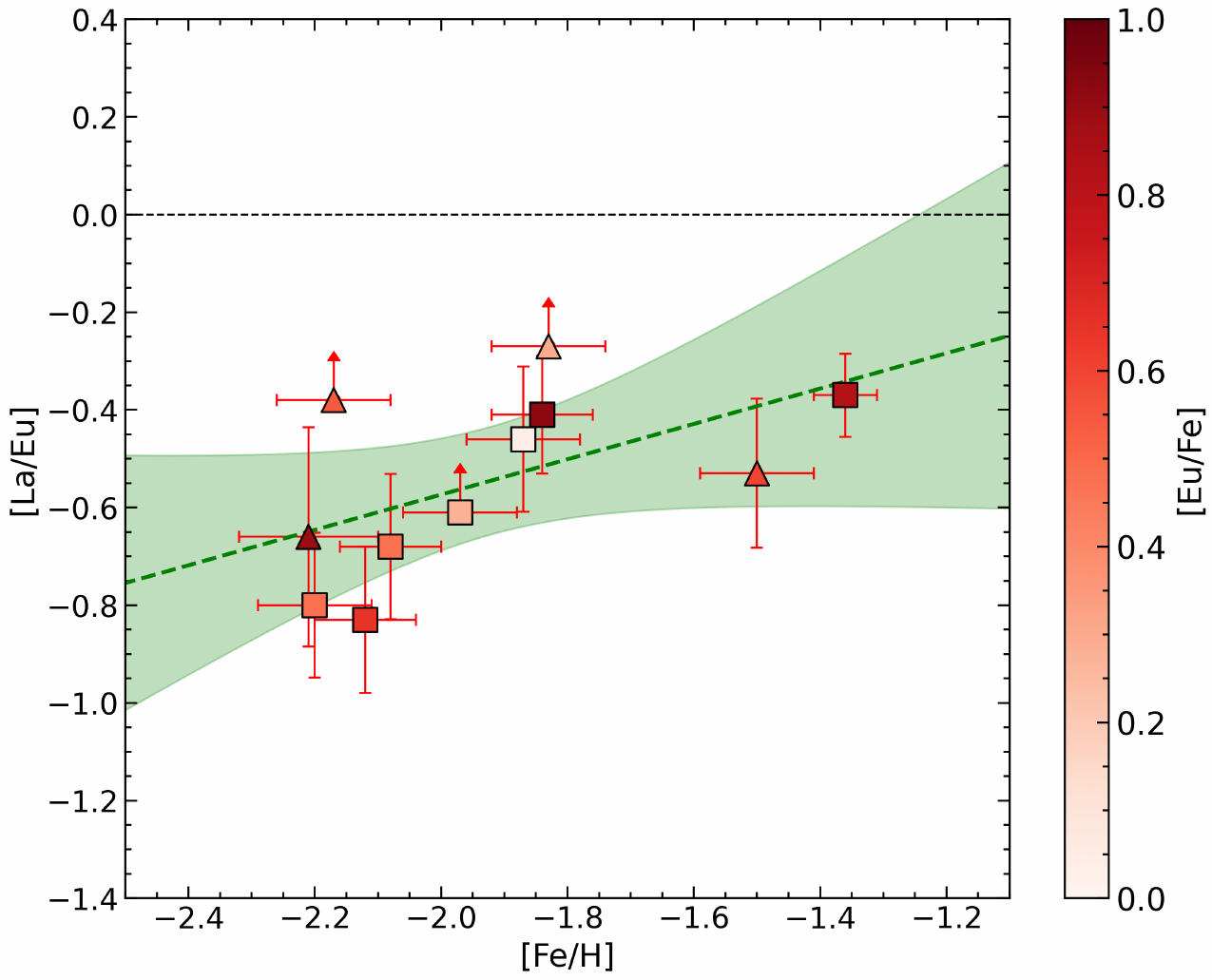}
        \includegraphics[width=0.40\textwidth]{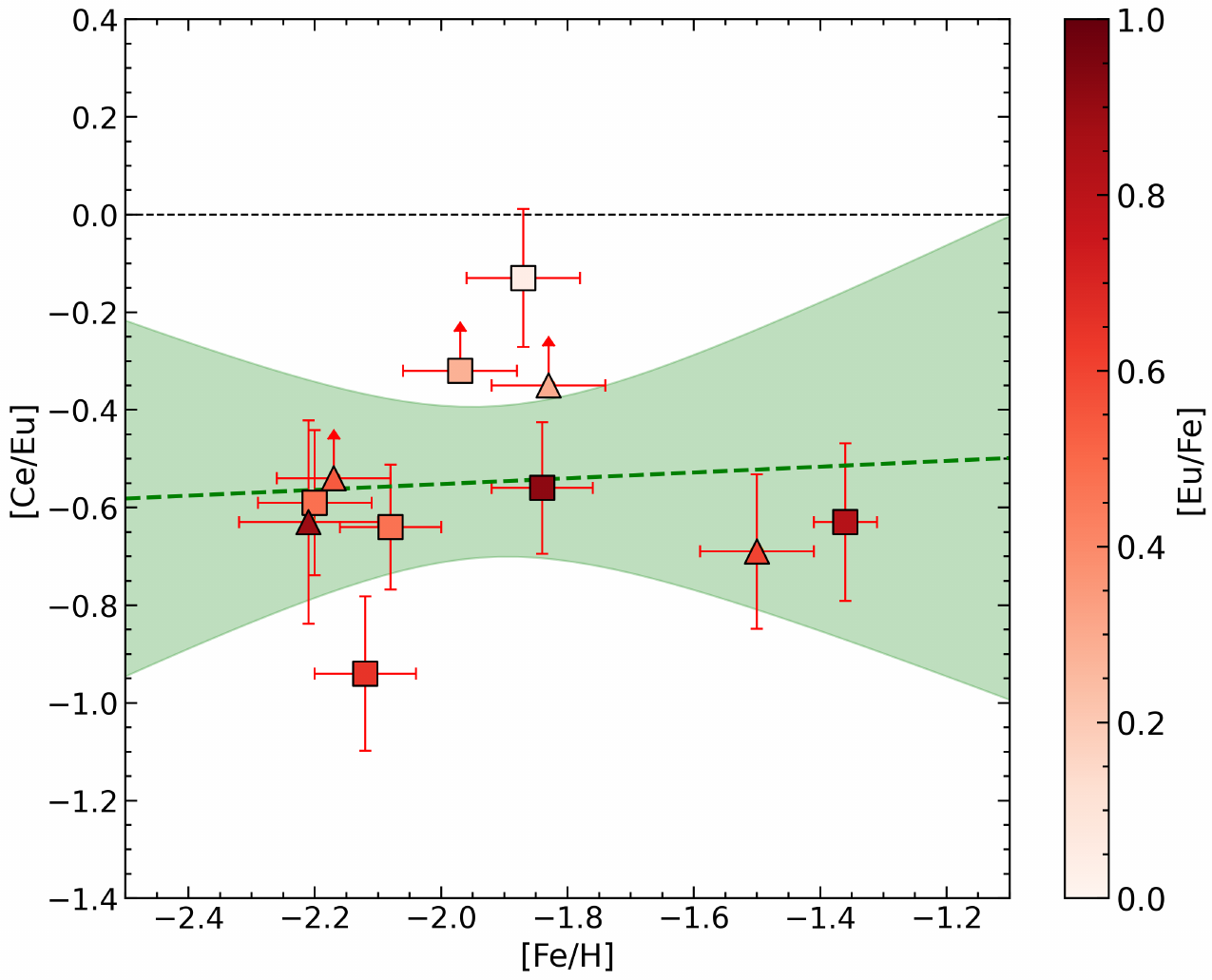}
        \includegraphics[width=0.40\textwidth]{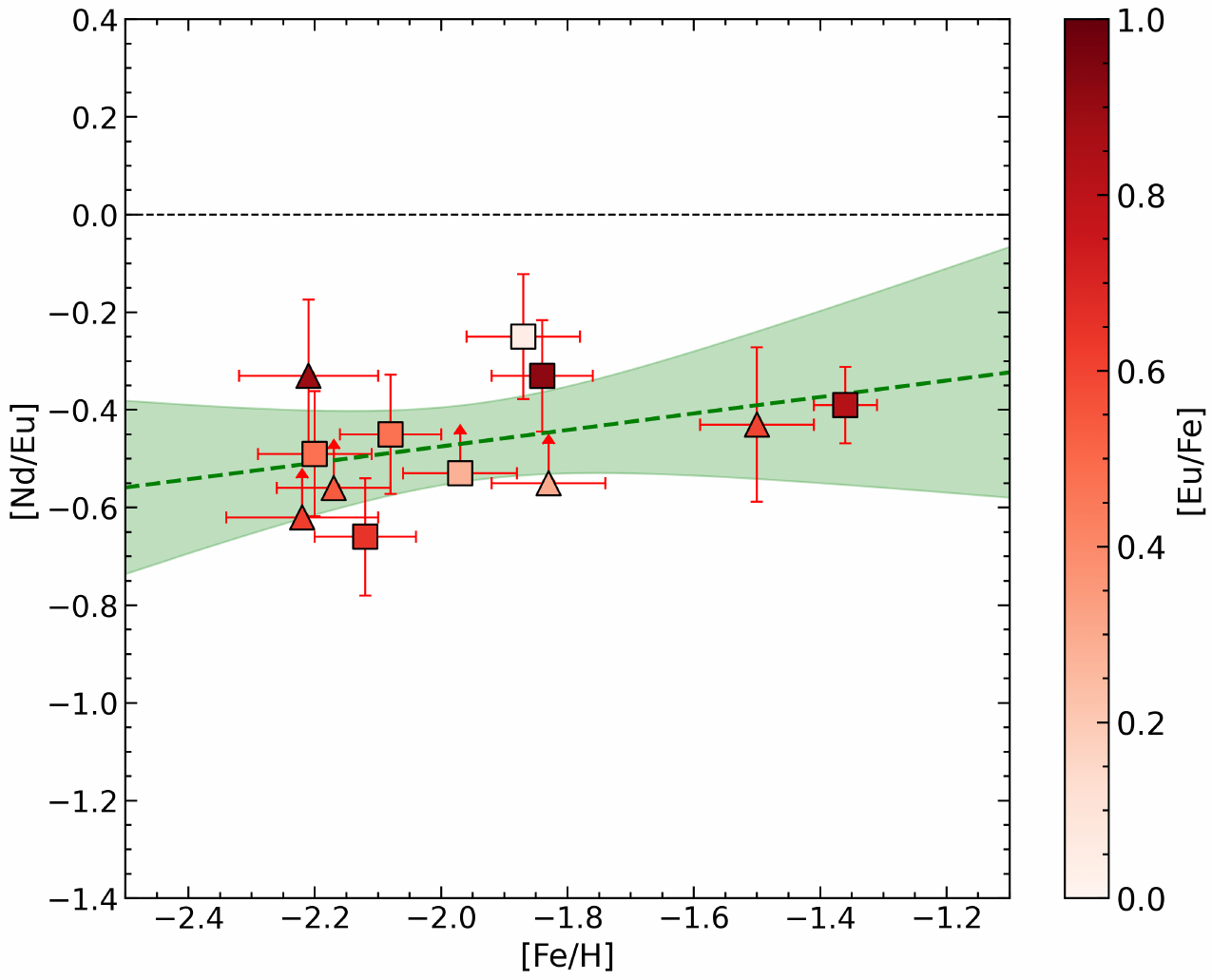}
        \includegraphics[width=0.40\textwidth]{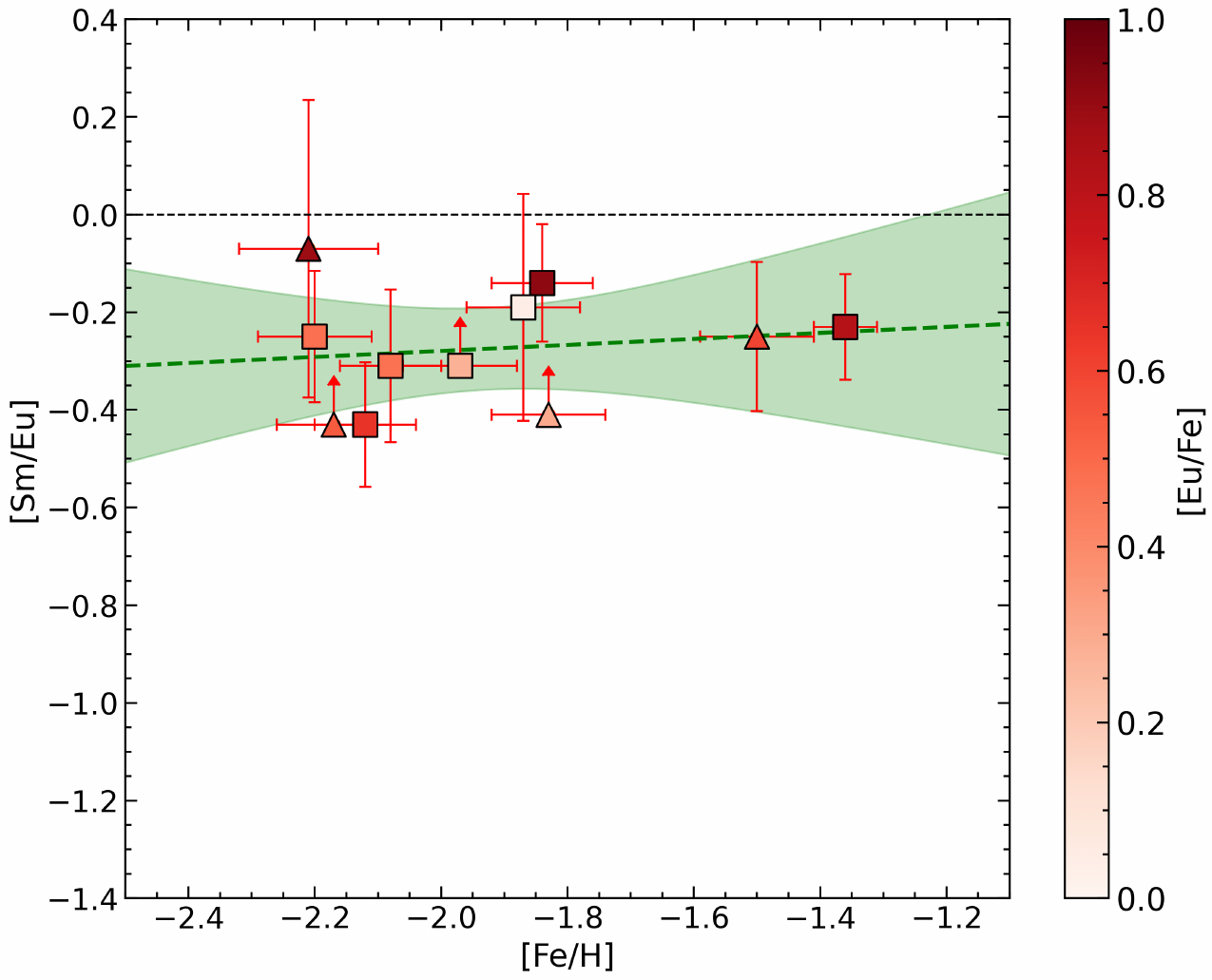}
        \caption{[Y/Eu], [Ba/Eu], [La/Eu], [Ce/Eu], [Nd/Eu] and [Sm/Eu] abundances as a function of [Fe/H] for SMC stars observed with UVES (squares) and MIKE (triangles). Points are coloured according to their [Eu/Fe] abundances. Lower limits are reported with arrows. In each plot, the linear regression of the measured abundances and their 95\% confidence intervals are shown with a green dashed line and green shaded area, respectively.}
        \label{sEu}
    \end{figure*}

    As pointed out in Sect.~\ref{iron}, [Fe/H] suggest most of the observed stars are likely $>$12 Gyr old according to the SMC AMR models. In particular, stars with [Fe/H]$<-1.8$ stars must have formed in the first Gyr of life of the SMC. These stars show mildly enhanced [$\alpha$/Fe] ratios indicating a major contribution by CC-SNe. In fact, the main site of production of $\alpha$-elements are massive stars ending their evolution after few tens million years with CC-SNe explosion, while there is just a marginal contribution from SNe Ia explosions. The delay between the enrichment timescales of CC-SNe and SNe Ia explains the behaviour of [$\alpha$/Fe] as a function of [Fe/H] \citep{Tinsley79, Matteucci86}. However, the values of [$\alpha$/Fe] in these stars are on average lower than those usually measured in MW stars of similar [Fe/H], as already pointed out by \citet{Nidever20}, \citet{Hasselquist21} and Paper I. This suggests a lower contribution to the early chemical enrichment of the SMC by massive stars and likely a $\alpha$-knee occurring at lower metallicities. This finding is in line with the lower star formation rate (SFR) of the SMC with respect to the MW \citep{Cignoni12,Cignoni13,Rubele18,Massana22}.

    By that time (in the first Gyr of the SMC life), most low-mass stars could not have evolved up to AGB phase and could not have already enriched the SMC gas with their $s$-process signature. Conversely to what happens at higher metallicities, i.e. at later times in the chemical evolution of a galaxy, the so called $s$-process elements actually lack their main channel of formation down to [Fe/H] $\sim-2.0$, also considering that even the few AGB stars have metallicities for which $s$-process nucleosynthesis is disfavoured. Within this scenario, the $r$-processes eventually become the main channel of nucleosynthesis for all neutron-capture elements \citep{Truran81}, and we expect to see the patterns and features of $r$-process elements (Eu, Sm) in the "$s$-process" elements (Y, Ba, La, Ce, Nd) as well.

    Indeed, if we compare the abundances of [Eu/Fe] or [Sm/Fe] (Figs.~\ref{Eu} and \ref{Sm}) with any of the observed $s$-process elements (Figs.~\ref{Y}-\ref{Nd}), we note that all these abundances share a similar $\sim1$ dex wide scatter in [X/Fe]. The large star-to-star [Eu/Fe] scatter among metal-poor Galactic stars (and in a similar way in the SMC) is caused by the rarity of events like NSM producing $r$-process elements. On the other hand, the metal-richer decrease of [Eu/Fe] is the result of the same delayed chemical contribution from SNe Ia at the basis of the decline of [$\alpha$/Fe]: in fact, SNe Ia are not expected to produce $r$-process elements, while they mainly inject iron-peak elements in the gas. 
     
    Moreover, as highlighted for multiple elements in Sect. \ref{s},  whenever a star has high [Eu/Fe] (or also [Sm/Fe]), it also shows high values of [$s$/Fe]. A comparison between [$s$/Eu] and [Fe/H] better visualizes the previous statement. In Fig.~\ref{sEu}, it is possible to see that, independently of the metallicity of our metal-poor stars, [$s$/Eu] abundances stand at a fairly constant level despite their scatter, very similarly to [Sm/Eu]. $s$-elements in the SMC have been produced at a constant ratio with Eu/$r$-process elements throughout the $-2.3<$ [Fe/H] $<-1.4$ range and they show consistently (and often markedly) subsolar values. These two features are clearly showing the $r$-processes are driving the nucleosynthesis of all the analysed neutron capture elements, while suggesting no relevant $s$-process acting at this metallicities, with only possible small $s$-process contribution from rotating massive stars. 

\section{Comparison with a stochastic chemical evolution model for the SMC}
\label{ss:stoc_CEM}

    To test our claims on more theoretical grounds, in the following we adopt a stochastic chemical evolution model to describe the early chemical evolution in the SMC. 

    The model resemble the scheme adopted in the stochastic models presented in \citet{Cescutti08,Cescutti10} and later adopted in other works to model the early evolution of the Galactic halo. Stochasticity is introduced by dividing the galaxy into isolated cubic regions, each containing the typical mass of gas swept by a CC-SNe (see \citealt{Cescutti08}). Within each cubic region, the model considers at each step gas inflow and outflow, gas recycling due to evolving stars and star formation. For the latter, masses of the stars formed are randomly extracted weighting them according to the initial mass function (here \citealt{Scalo86}). More details on the physical prescriptions adopted in the model are given in Appendix \ref{aa:stoc_CEM}. However, it is worth to point out that to model the chemical evolution of the SMC we adopt a smaller value of the star formation efficiency (SFE) to what is routinely adopted for the MW halo, namely of $\nu=0.25\ {\rm Gyr^{-1}}$. The latter value is in line with values usually adopted to model the evolution of the MCs (see, e.g. \citealt{Vasini23}). Adopting this prescription for the SF, we sample a number of 100 volumes, which results in a summed total mass at the end of the run (of 1.5 Gyr) of $\sim2\times 10^7 {\rm \textit{M}_\odot}$, similar to the stellar mass inferred from CMD fitting in the same age range by \citet[][$\sim5\times 10^7 {\rm \textit{M}_\odot}$]{Massana22}.

    To trace the evolution of neutron-capture elements, the model includes different sources for both $s$-process and $r$-process synthesis. $S$-process is synthesised in rotating massive stars and AGB stars, for which we adopt the stellar yields by \citet[][adopting the rotational velocity distribution by \citealt{Rizzuti21}]{Limongi18} and the FRUITY database (e.g. \citealt{Cristallo11}), respectively. $R$-process is instead competitively produced in MRD-SNe and NSM in the model. For MRD-SNe, we adopt the yields of \citet{Nishimura17} and a fraction in the progenitor mass range  of 0.1. For NSM, we adopt the same yields as in \citet[][derived from Sr measurement in the Kilonova AT2017gfo, \citealt{Watson19}]{Molero23}, a binary fraction in the progenitor range of 0.08 (as derived in \citealt{Palla25}) and a fixed coalescence timescale of 100 Myr. This timescale is set to distinguish clearly the effect between the prompt (MRD-SNe) and the delayed (NSM) source of $r$-process production. For further details on the nucleosynthesis prescriptions, we address the reader to Appendix \ref{aa:stoc_CEM}.

    \begin{figure}
        \centering
        \includegraphics[width=0.475\textwidth]{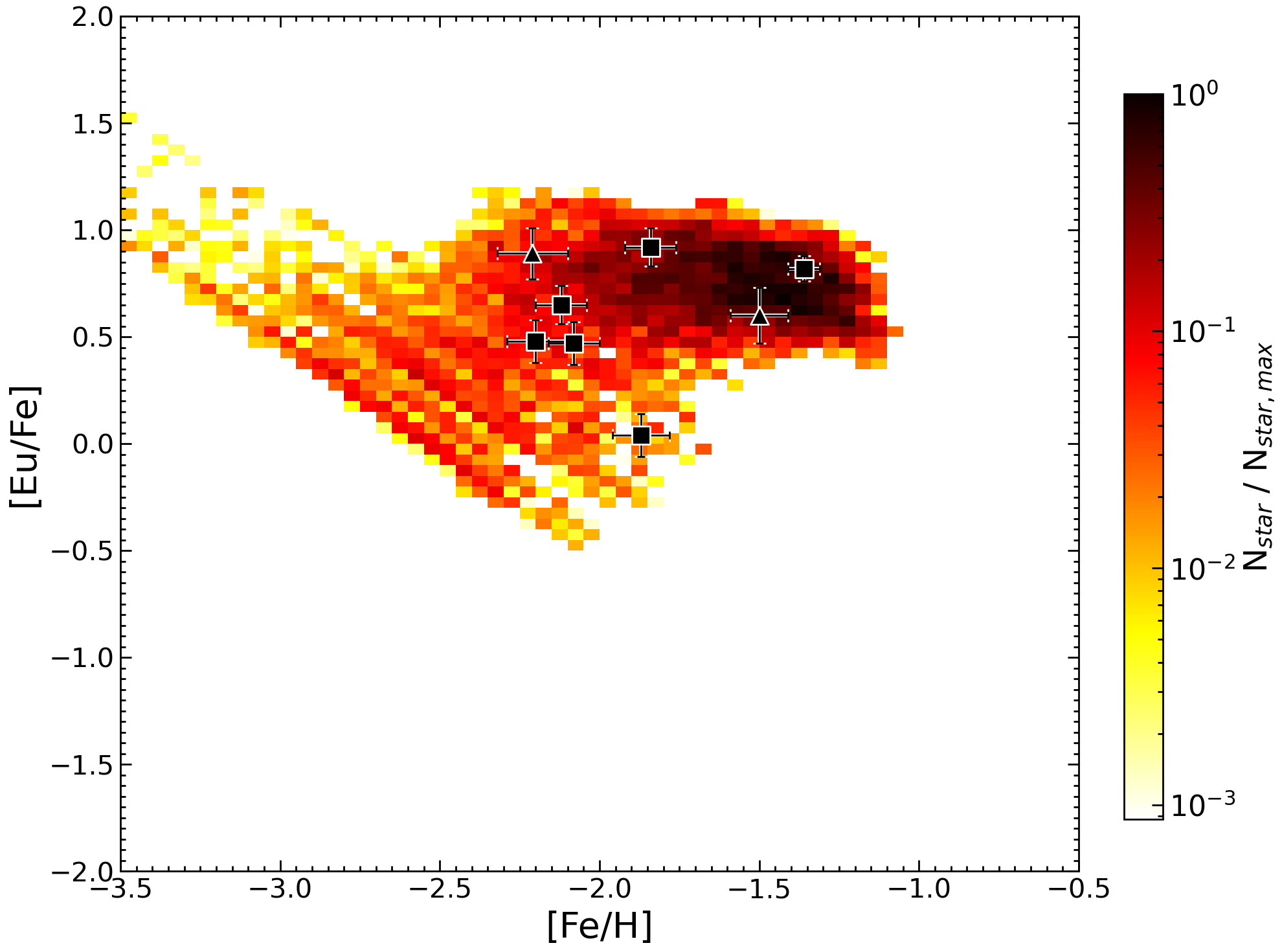}
        \caption{[Eu/Fe] abundances as a function of [Fe/H] as predicted by our adopted stochastic chemical evolution model. The colormap displays the number density of predicted long-lived stars by the model on a logarithmic scale. Observed SMC abundances with relative errorbar are represented with black symbols (squares: UVES observations, triangles: MIKE observations).}
        \label{Eu_stochastic}
    \end{figure}

    In Fig. \ref{Eu_stochastic}, we show the output of our fiducial stochastic chemical evolution model with the [Eu/Fe] abundances derived for metal-poor SMC stars. Observations are well reproduced by the stochastic model, which also reflects the abundance spread observed in stars in our sample. In turn, this confirm our indication that such a spread of $\sim1$ dex in [Eu/Fe] is a consequence of inhomogeneous mixing in the early evolution of the SMC, in analogy to what is observed in the MW at similar metallicities (see Fig. \ref{Eu}). However, to achieve the result in Fig. \ref{Eu_stochastic}, in the SMC model we have to increase significantly the fraction of system originating NSM relative to Galactic studies. Indeed the adopted fraction for the MW span a range between $2\times 10^{-3}$ and $0.02$ (e.g. \citealt{Molero21,Molero23,Cavallo21}), whereas here we adopt 0.08, according to the hypothesis by \citet{Palla25} of a larger fraction of NSM progenitor to reproduce MW satellite europium patterns at low-metallicity. It is worth noting that to probe the effect of NSM progenitor fraction in the early SMC evolution, we run an additional model with lower fraction of NSM progenitor ($6 \times 10^{-3}$, in the range of values adopted for the MW). The comparison between the two models is shown in Fig. \ref{Eu_compare} and shows clearly the need for a large fraction of NSM at low-metallicity to reproduce the upper envelope of [Eu/Fe] abundances, indicating a higher relative rate of $r$-process events in the Cloud. 

    \begin{figure*}
        \centering
        \includegraphics[width=0.475\textwidth]{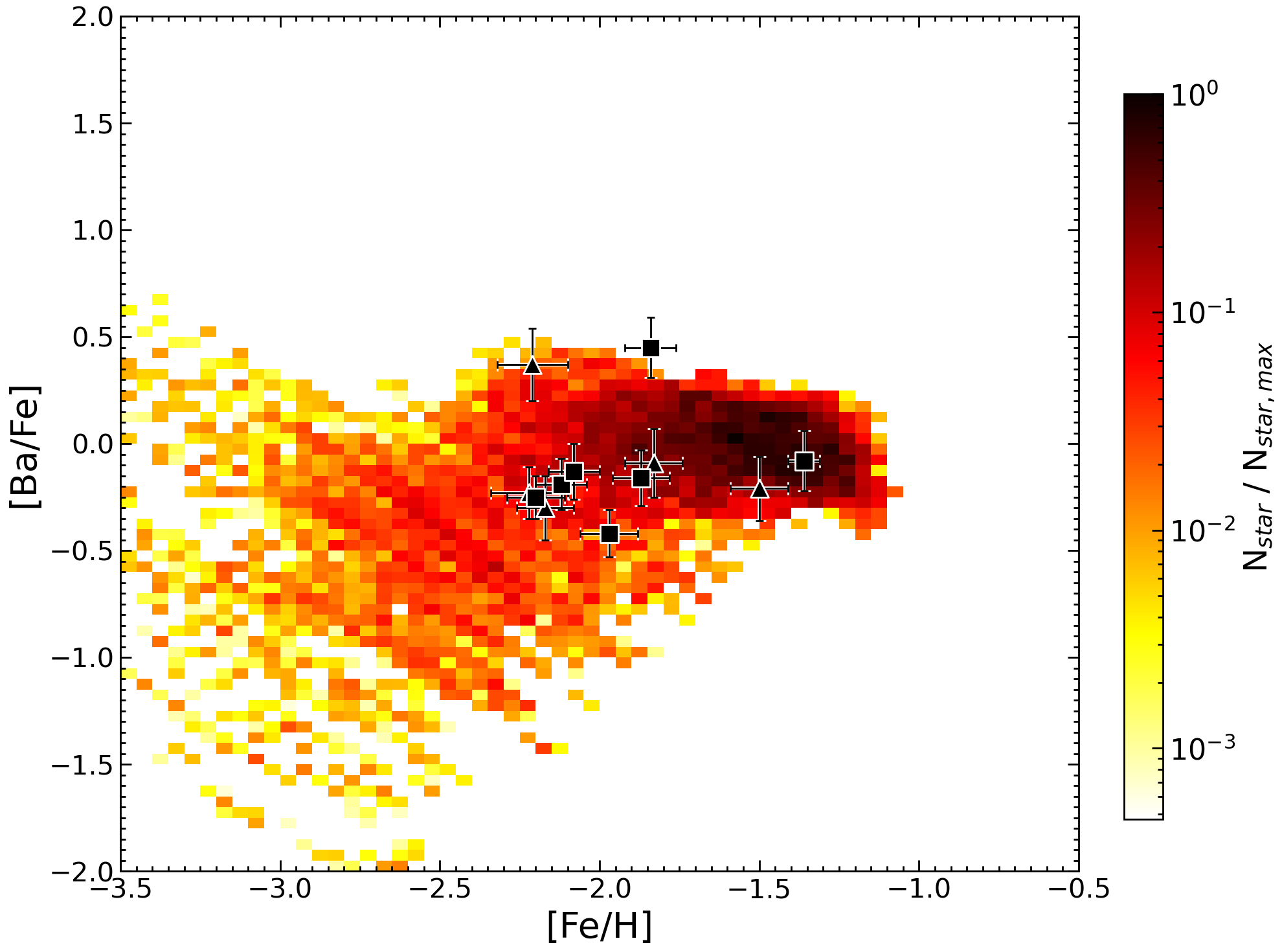}
        \includegraphics[width=0.475\textwidth]{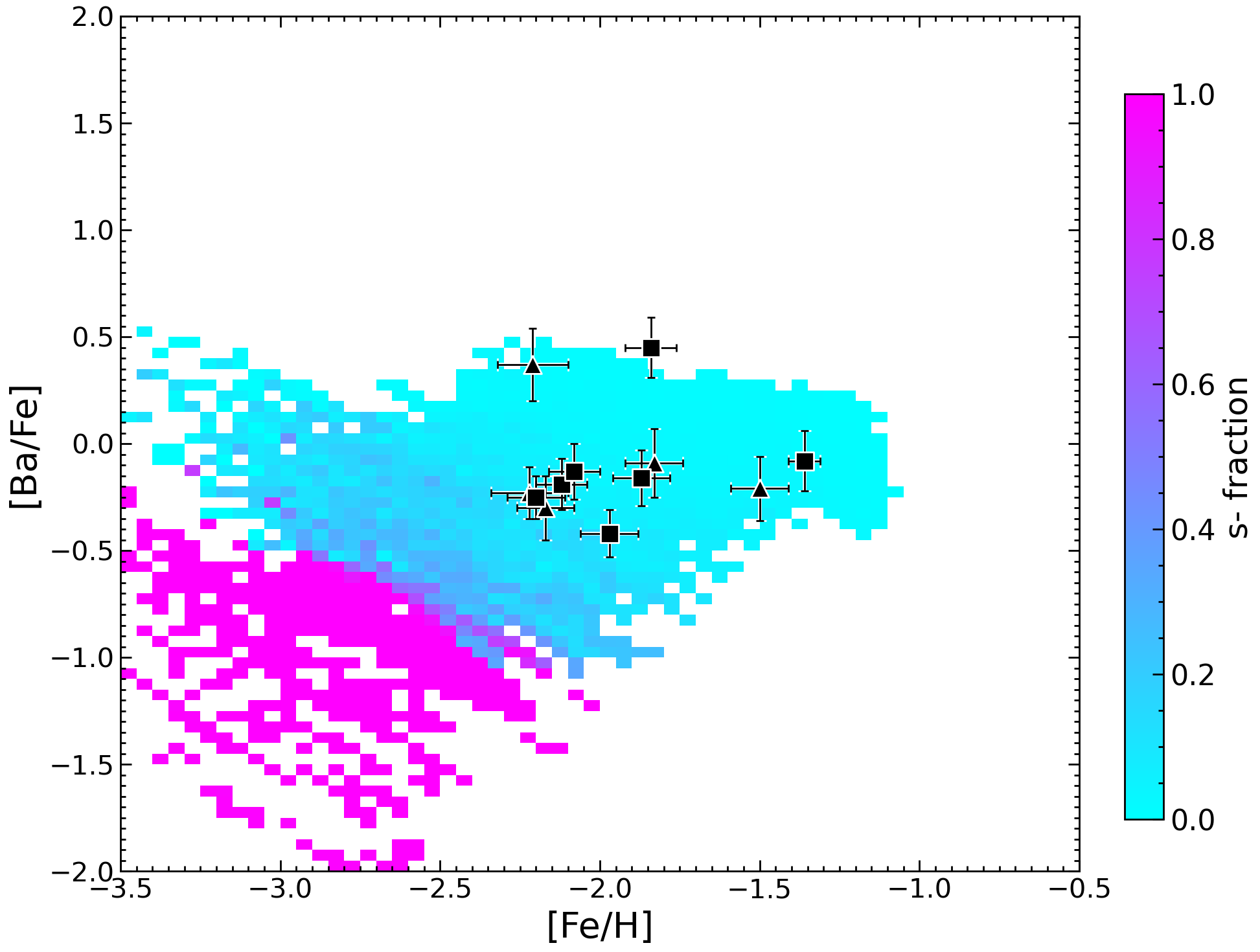}
        \caption{[Ba/Fe] abundances as a function of [Fe/H] as predicted by our adopted stochastic chemical evolution model. On the left panel, the colormap displays the number density of predicted long-lived stars by the model on a logarithmic scale. On the right panel, the colormap displays the fraction of element abundance produced by s-process over total. Black symbols are as in Fig. \ref{Eu_stochastic}.}
        \label{Ba_stochastic}
    \end{figure*}

    In Fig. \ref{Ba_stochastic}, we show instead the comparison between our fiducial stochastic chemical evolution model and SMC stars for [Ba/Fe], being Ba a typical $s$-process element. In particular, Fig. \ref{Ba_stochastic} left panel shows the predicted density of long-lived stars, while right panel displays the fraction of Ba abundance produced by the $s$-process (both from AGBs and rotating massive stars). The two panels show clearly that in order reproduce the pattern and spread of SMC metal-poor stars we need that the majority of Ba production comes from the $r$-process: low-mass AGB stars do not have sufficient time to pollute the ISM at these early stages of evolution, while rotating massive stars are responsible of the low-metallicity, [Ba/Fe]-poor end of the predictions, where no data are encountered. This strikingly highlights the claim of $r$-process driven production for neutron-capture elements in the SMC brought out  in Sect. \ref{ss:rproc_sproc}. Moreover, further tests on enhanced/more significant $s$-process production from rotating massive stars reveal this scenario to be unlikely. Indeed, adopting a rotational velocity distribution more in favour of rapidly rotating stars relative to the fiducial one (see Figure \ref{Ba_sfrac}) leads to an extreme enhancement in Ba ($\gtrsim$0.5 dex at [Fe/H]$\sim-1.5$ dex), which is not followed by the data trend.

\section{Conclusions} \label{conclusions}

    We analysed high resolution  UVES and MIKE spectra for 12 metal-poor SMC RGB stars focusing on their abundances of neutron-capture elements 
    (in particular the pure $r$-process elements Eu and Sm).
    The main results emerging from this chemical analysis are summarised in the following:

    \begin{itemize}
        \item the targets have [Fe/H] between $-2.3$ and $-1.4$ dex, 10 out 12 with [Fe/H]$<-1.8$ dex; according to the theoretical AMR by \citet{Pagel98}, these stars formed in the first Gyr of life of the SMC;
        \item $\alpha$-elements are enhanced in metal-poor SMC stars, but at a lower level with respect to MW stars of similar [Fe/H]. This lower [$\alpha$/Fe] enhancement and a mild decrease of [$\alpha$/Fe] for increasing metallicity are consistent with the SFR of the SMC, less efficient than that of the MW. This dataset suggests that the $\alpha$-knee occurs at lower [Fe/H] than in the MW, but it is not yet possible to properly identify it. This is visible in Fig.~\ref{alpha}: metal-poor SMC stars draw a decreasing trend reaching solar or occasionally mildly subsolar [$\alpha$/Fe] both in hydrostatic (Mg and O) and in explosive (Si and Ca) $\alpha$-elements;
        \item the targets cover a large range of [Eu/Fe], from solar value up to $\sim+1$ dex, with three stars showing [Eu/Fe]$>+0.7$. This large star-to-star scatter, not explainable within the uncertainties, is similar to that observed in metal-poor MW stars. Analogue pattern is observed for [Sm/Fe]. Such pattern indicates that the $r$-process in the SMC can be extremely efficient (as claimed in R21), but still largely affected by the stochastic nature of the main sites of production and inefficient gas mixing in the early SMC evolution. These findings are in line with predictions from a stochastic chemical evolution model simulating the early evolution of the SMC; 
        \item also the $s$-process elements show large star-to-star scatter not compatible within the uncertainties. Whenever a star has high [Eu/Fe], it also shows high [$s$/Fe]. Moreover, we observe all stars have markedly subsolar [$s$/Eu]. At these metallicities, the nucleosynthesis of the neutron capture elements is driven by $r$-processes, since the low-mass AGB stars have not yet evolved and left their $s$-process signature in the ISM. This interpretation is confirmed by results of the stochastic chemical evolution model, also suggesting that $s$-process production in rotating massive stars is only accounting for a very minor fraction of the abundance in the observed stars.
    \end{itemize}

    Summarizing, the SMC metal-poor stars display distinct [$\alpha$/Fe] with respect to MW metal-poor stars. Indeed, the $\alpha$ patterns suggest the SMC has a lower SFR than the MW, leading to a lower contribution by massive stars to the overall chemical enrichment of the galaxy \citep{Matteucci12}. On the other hand, all neutron capture elements, including those usually identified as $s$-process, are mostly driven by $r$-process production, exhibiting the typical characteristics (scatter and relative abundances) of $r$-process nucleosynthesis.

\begin{acknowledgements}
A.M., M.P. and D.R. acknowledge support from the project "LEGO – Reconstructing the building blocks of the Galaxy by chemical tagging" (P.I. A. Mucciarelli) granted by the Italian MUR through contract PRIN 2022LLP8TK\_001. L.M. gratefully acknowledges support from ANID-FONDECYT Regular Project n. 1251809. D.A.A.G. acknowledges funding from the European Union under the grant ERC-2022-AdG, ‘StarDance: the non-canonical evolution of stars in clusters’, Grant Agreement 101093572, PI: E. Pancino.

\end{acknowledgements}

\bibliography{references.bib}

\appendix

\section{Linelists and \texorpdfstring{$\log gf$}{} references} 
\label{atomic_data}

    We summarise in Table \ref{loggf} the references for the $\log gf$ values we used in the analysis of our targets.

    \begin{table}[!ht]
        \caption{Literature references for $\log gf$ values of the analysed elements.}
        \label{loggf}
        \centering
        \begin{tabular}{c c}
        \hline\hline
        Element & $\log gf$ reference(s) \\
        \hline 
        [O I] & \cite{Wiese96} \\
        Mg I & \cite{PehlivanRhodin17} \\
        Si I & \cite{DenHartog23} \\
        Ca I & \cite{DenHartog21} \\
        Fe I & \cite{Fuhr88, Fuhr06} \\
        & \cite{Ruffoni14, DenHartog14} \\
        Fe II & \cite{Melendez09} \\
        Y II & \cite{Biemont11} \\
        Ba II & \cite{Wiese80,Klose02} \\
        La II & \cite{Lawler01a} \\
        Ce II & \cite{Lawler09} \\
        Nd II & \cite{DenHartog03} \\
        Sm II & \cite{Lawler06} \\
        Eu II & \cite{Lawler01b} \\
        \hline
        \end{tabular}
    \end{table}

\section{Additional information on chemical abundance ratios of SMC metal-poor stars}\label{appendix_b}

    In this Section, we discuss more in detail the determination of C and N abundances and their measured values. Finally, in Table \ref{aburatios} we provide the full set of [X/Fe] abundances measured from our sample of SMC metal-poor targets and discussed in Sect. \ref{abundances}.
    
    The abundances of C, N and O are interconnected each other through the molecular equilibrium, therefore each abundance is sensitive to the abundances of the other two elements \citep[see e.g.][]{ryde09}. For these elements, we adopted an iterative scheme. Adopting the O abundances derived from the forbidden O line at 6300.3 \AA\ (see Section~\ref{alfa}) and a solar-scaled N abundance, C abundances were derived from the G-band, then fixing [C/Fe], N abundances were obtained from CN bands, and O abundances are newly determined assuming the new C and N values. New abundances for the three elements are re-obtained until the abundance variations between the consecutive iterations were smaller than 0.05 dex.

    All the UVES targets have [C/Fe]$\sim-1.0$ dex and [N/Fe]$\sim+0.5$ dex, as expected for stars experiencing the mixing episode at the RGB Bump level \citep{Gratton00, Spite05}, therefore these abundances do not reflect the pristine C and N abundances of these stars that are modified by additional effects, i.e. thermohaline mixing  \citep{lagarde12}. Despite these abundance modifications, [C+N/Fe] should be conserved \citep{salaris15}. The measured [C+N/Fe] abundances (Table \ref{aburatios}) range from $-0.5$ and $+0.0$ dex that are fully compatible with abundances measured by APOGEE in SMC stars with [Fe/H] $\sim-2$ dex \citep{Hasselquist21}. Finally, we note that none of the stars shows high carbon values ([C/Fe] $>+1.0$ dex), therefore none of them can be considered carbon-enhanced metal-poor stars.

    \begin{table*}
        \caption{[X/Fe] chemical abundances of the observed SMC stars.}
        \label{aburatios}
        \centering
        \begin{tabular}{c c c c c c c c}
        \hline\hline
        & UVES-1 & UVES-2 & UVES-3 & UVES-4 & UVES-5 & UVES-6 & UVES-7 \\
        \hline 
        \text[C/Fe]  & $-0.64\pm0.12$ & $-1.20\pm0.12$ & $-1.00\pm0.12$ & $-1.07\pm0.12$ & $-1.15\pm0.12$ & $-0.78\pm0.12$ & $-0.93\pm0.12$ \\
        \text[N/Fe]  & $+0.50\pm0.08$ & $+0.80\pm0.08$ & $+0.50\pm0.08$ & $+0.44\pm0.08$ & $+0.41\pm0.08$ & $+0.26\pm0.08$ & $+0.52\pm0.08$ \\
        \text[C+N/Fe]& $-0.16\pm0.09$ & $+0.02\pm0.08$ & $-0.22\pm0.08$ & $-0.26\pm0.08$ & $-0.34\pm0.09$ & $-0.18\pm0.09$ & $-0.51\pm0.09$ \\
        \text[O/Fe]  & $+0.53\pm0.09$ & $+0.67\pm0.09$ & $+0.50\pm0.08$ & $+0.27\pm0.12$ & $+0.53\pm0.10$ & $+0.22\pm0.06$ & $+0.29\pm0.09$ \\
        \text[Mg/Fe] & $-0.10\pm0.04$ & $+0.29\pm0.07$ & $+0.11\pm0.11$ & $+0.20\pm0.06$ & $+0.15\pm0.06$ & $+0.22\pm0.06$ & $+0.13\pm0.06$ \\
        \text[Si/Fe] & $+0.11\pm0.08$ & $+0.49\pm0.09$ & $+0.23\pm0.11$ & $+0.18\pm0.09$ & $+0.30\pm0.07$ & $+0.24\pm0.08$ & $+0.25\pm0.08$ \\
        \text[Ca/Fe] & $+0.07\pm0.03$ & $+0.30\pm0.03$ & $+0.25\pm0.03$ & $+0.21\pm0.03$ & $+0.22\pm0.04$ & $+0.28\pm0.15$ & $+0.11\pm0.02$ \\
        \text[Y/Fe]  & $-0.09\pm0.09$ & $-0.62\pm0.09$ & $-0.41\pm0.08$ & $-0.70\pm0.08$ & $-0.56\pm0.08$ & $-0.18\pm0.14$ & $-0.81\pm0.09$ \\
        \text[Ba/Fe] & $+0.45\pm0.14$ & $-0.25\pm0.10$ & $-0.19\pm0.12$ & $-0.16\pm0.13$ & $-0.13\pm0.13$ & $-0.08\pm0.14$ & $-0.42\pm0.11$ \\
        \text[La/Fe] & $+0.51\pm0.08$ & $-0.32\pm0.11$ & $-0.18\pm0.12$ & $-0.42\pm0.11$ & $-0.21\pm0.11$ & $+0.45\pm0.06$ & $-0.33\pm0.15$ \\
        \text[Ce/Fe] & $+0.36\pm0.10$ & $-0.11\pm0.11$ & $-0.29\pm0.13$ & $-0.09\pm0.10$ & $-0.17\pm0.08$ & $+0.19\pm0.15$ & $-0.04\pm0.12$ \\
        \text[Nd/Fe] & $+0.59\pm0.07$ & $-0.01\pm0.08$ & $-0.01\pm0.08$ & $-0.21\pm0.08$ & $+0.02\pm0.07$ & $+0.43\pm0.05$ & $-0.25\pm0.11$ \\
        \text[Sm/Fe] & $+0.78\pm0.08$ & $+0.23\pm0.09$ & $+0.22\pm0.09$ & $-0.15\pm0.21$ & $+0.16\pm0.12$ & $+0.59\pm0.09$ & $-0.03\pm0.11$ \\
        \text[Eu/Fe] & $+0.92\pm0.09$ & $+0.48\pm0.10$ & $+0.65\pm0.09$ & $+0.04\pm0.10$ & $+0.47\pm0.10$ & $+0.82\pm0.06$ & $<+0.28$ \\
        \hline
        \end{tabular}
        \begin{tabular}{c c c c c c}
        \hline\hline
        & MIKE-1 & MIKE-2 & MIKE-3 & MIKE-4 & MIKE-5 \\
        \hline
        \text[O/Fe]  & $...$ & $...$ & $...$ & $+0.41\pm0.12$ & $...$ \\
        \text[Mg/Fe] & $+0.13\pm0.11$ & $+0.26\pm0.10$ & $+0.18\pm0.12$ & $-0.12\pm0.10$ & $0.00\pm0.04$ \\
        \text[Si/Fe] & $+0.22\pm0.14$ & $+0.11\pm0.08$ & $+0.29\pm0.15$ & $-0.02\pm0.11$ & $-0.08\pm0.09$ \\
        \text[Ca/Fe] & $+0.18\pm0.05$ & $+0.24\pm0.04$ & $+0.30\pm0.07$ & $-0.06\pm0.05$ & $+0.09\pm0.04$ \\
        \text[Y/Fe]  & $-0.20\pm0.18$ & $-0.39\pm0.11$ & $...$ & $-0.81\pm0.15$ & $-0.51\pm0.12$ \\
        \text[Ba/Fe] & $+0.37\pm0.17$ & $-0.30\pm0.15$ & $-0.23\pm0.12$ & $-0.09\pm0.16$ & $-0.21\pm0.15$ \\
        \text[La/Fe] & $+0.23\pm0.19$ & $+0.16\pm0.15$ & $...$ & $+0.03\pm0.16$ & $+0.07\pm0.08$ \\
        \text[Ce/Fe] & $+0.26\pm0.17$ & $0.00\pm0.11$ & $...$ & $-0.05\pm0.17$ & $-0.09\pm0.09$ \\
        \text[Nd/Fe] & $+0.56\pm0.10$ & $-0.02\pm0.08$ & $0.00\pm0.17$ & $-0.25\pm0.07$ & $+0.17\pm0.09$ \\
        \text[Sm/Fe] & $+0.82\pm0.28$ & $+0.11\pm0.17$ & $...$ & $-0.11\pm0.10$ & $+0.35\pm0.08$ \\
        \text[Eu/Fe] & $+0.89\pm0.12$ & $<+0.54$ & $<+0.62$ & $<+0.30$ & $+0.60\pm0.13$ \\
        \hline
        \end{tabular}
    \end{table*}

\section{Additional details and runs for the stochastic SMC model}

In this Section, we provide more details on the assumptions and physical prescriptions of the model presented in Section \ref{ss:stoc_CEM} (\ref{aa:stoc_CEM}), together with test of the model output onto $\alpha$-elements (\ref{aa:alpha}) and results for additional model runs on neutron-capture elements (\ref{aa:ncapt}).

\subsection{Model prescriptions}
\label{aa:stoc_CEM}

    The dimensions of the independent cubic regions of the model is chosen in order to neglect the interaction between different regions (at least in first approximation). For typical ISM densities, a SN remnant becomes indistinguishable from the ISM – i.e. merges with the ISM – before reaching $\sim50$ pc \citep{Thorton98}. Therefore, we adopt volumes with surface area of 2$\times 10^4$ pc$^{2}$ to ensure at the same time a good level of stochasticity, as larger volumes will provide more homogeneous results. For every volume, we adopt an evolutionary timestep of 1 Myr. In this way, we are able to trace precisely every stellar death, since the minimum stellar lifetime (see, e.g. \citealt{Schaller92,Gibson97}) is of $\sim3$ Myr. With this timestep, we are also able to assume the instantaneous mixing approximation, as cooling timescales that allow mixing between stellar ejecta are typically smaller (\citealt{Cescutti08} and references therein).
    
    We compute the chemical evolution in each region by using the usual chemical evolution equation, namely:
    \begin{equation}
        \dot{G}_i(t) =  \psi(t)\ X_i(t) + R_i(t) + \dot{G}_{i;inf} (t)  - \dot{G}_{i;out}(t),  
    \end{equation}
    where $G_i$ is the mass fraction of the element $i$ in the gas and $X_i$ represents the abundance in mass of the element \citep{Matteucci12}. The term $\dot{G}_{i;inf}$ accounts for the infall of gas, assumed to have primordial composition and following an exponential law:
    \begin{equation}
        \dot{G}_{inf} (t) = A\ e^{-t/\tau_{inf}},
    \end{equation}
    where $A$ is the infall normalization constant and $\tau_{inf}$ the infall timescale, in our case assumed to be 0.5 Gyr. SFR is parametrised according to the Schmidt-Kennicutt law \citep{Kennicutt98}:
    \begin{equation}
        \psi(t) = \nu \ \Sigma_{gas}(t)^{1.5},  
    \end{equation}
    where $\Sigma_{gas}$ is the gas mass surface density and $\nu$ is the SFE, set to 0.25 Gyr$^{-1}$ to model the SMC evolution. Moreover, the model takes an outflow from the system into account, proportional to the SFR:
    \begin{equation}
        \dot{G}_{out}(t) = \omega \ \psi(t),
    \end{equation}
    where $\omega$ is the mass loading factor, set to be 3 in our case. Finally, $R_i$ takes into account the ejecta from the different type of stars, including AGB, massive stars / CC-SNe, SNe Ia and exotic objects (as MRD-SNe and NSM). For standard sources, the yields are from the FRUITY database (\citealt{Cristallo11}, AGB stars), \citet[][CC-SNe]{Limongi18} and \citet[][SNe Ia]{Iwamoto99}.
 
    In the model, the stochasticity is taken into account in the masses of the new born stars. Indeed, at each timestep, in each volume, newborn star masses are randomly extracted weighting them according to the initial mass function (here from \citealt{Scalo86}). To allow the model to form at any timestep stars of any mass up to the considered maximum stellar mass, therefore preventing a bias toward low-mass stars, we impose a threshold in SFR of 100 \textit{M}$_\odot$. Events as SNe Ia, MRD-SNe and NSM are also randomly extracted from the IMF, taking into account the mass range of their progenitors  and the probability / fraction of event within the  mass range. In particular, for SNe Ia we consider a range 3-16 \textit{M}$_\odot$ and a probability of suitable binary systems of 0.05 (see \citealt{Romano05}), for MRD-SNe a mass range of 10-30 \textit{M}$_\odot$ and a fraction of 0.1 (see \citealt{Molero23}) and for NSM a mass range 8-50 \textit{M}$_\odot$ and a probability of 0.08 (see \citealt{Molero23,Palla25} for more details). Concerning the adopted delay-time-distributions (DTDs), for SNe Ia we follow the prescriptions of the single degenerate scenario of \citet{Matteucci86}, by randomly extracting the mass of the secondary (giving the clock to the system) according to the given distribution function (see \citealt{MR01}). For NSM, we instead adopt a fixed DTD of 100 Myr: the adoption of a fixed DTD instead of more complicated functional forms allows to identify more clearly the effects of these sources to the early chemical enrichment of the galaxy. Moreover, the 100 Myr delay is smaller than the typical timescale of SNe Ia, but large enough to distinguish these enrichment events from prompt MRD-SNe. 

\subsection{Reproducing the $\alpha$-elements}
\label{aa:alpha}

    To provide further justification that the stochastic chemical evolution model described in Section \ref{ss:stoc_CEM} and \ref{aa:stoc_CEM} represents a realistic picture of the early evolution of the SMC, in the following we display the comparison between the model and the metal-poor SMC stars for $\alpha$-elements. In fact, these elements have well defined nucleosynthetic origin (see, e.g. \citealt{Kobayashi20}) and are therefore less prone to significant uncertainties in their production, as in the case for neutron-capture elements. 

    \begin{figure}
        \centering
        \includegraphics[width=0.475\textwidth]{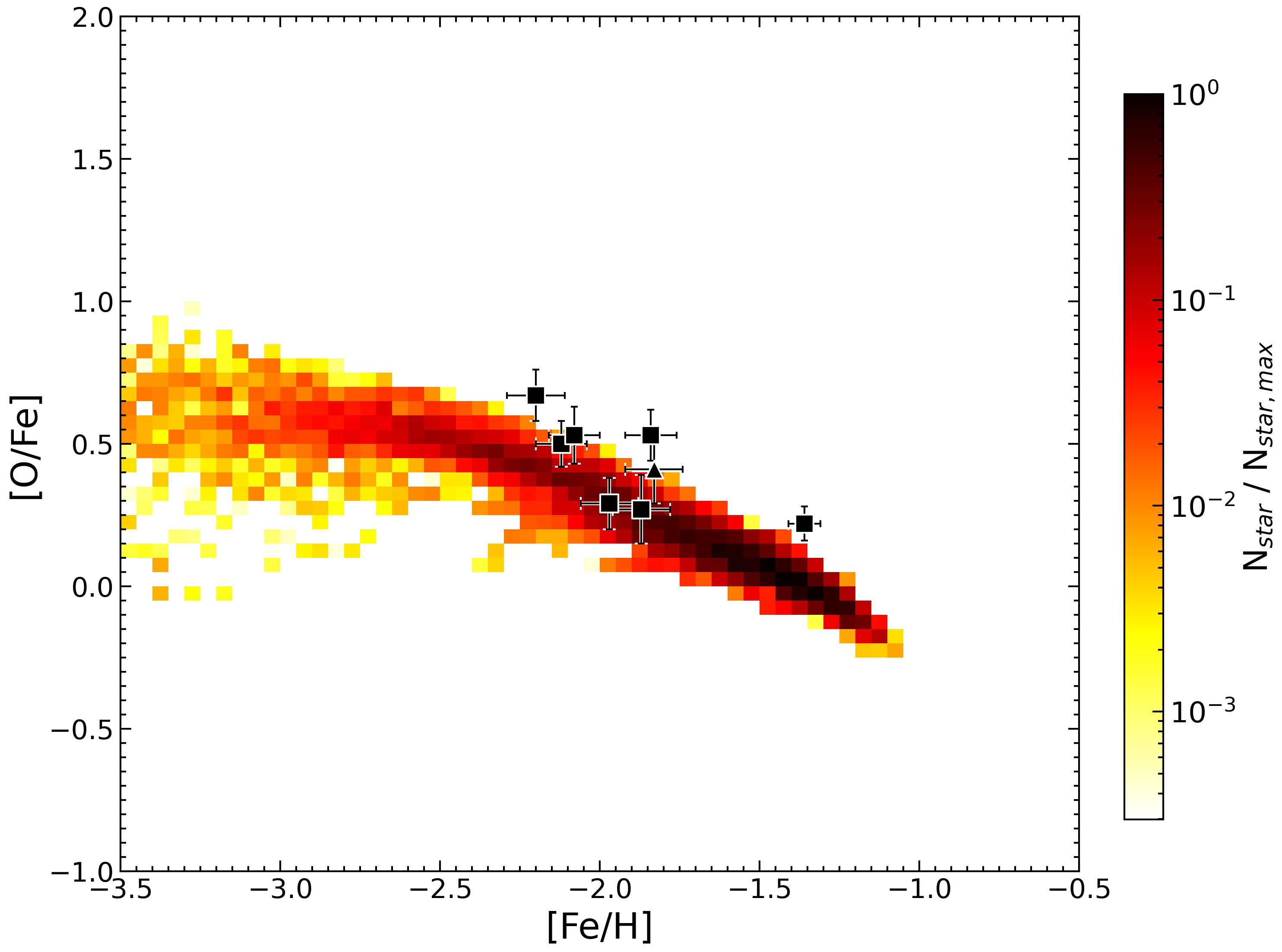}\\
        \includegraphics[width=0.475\textwidth]{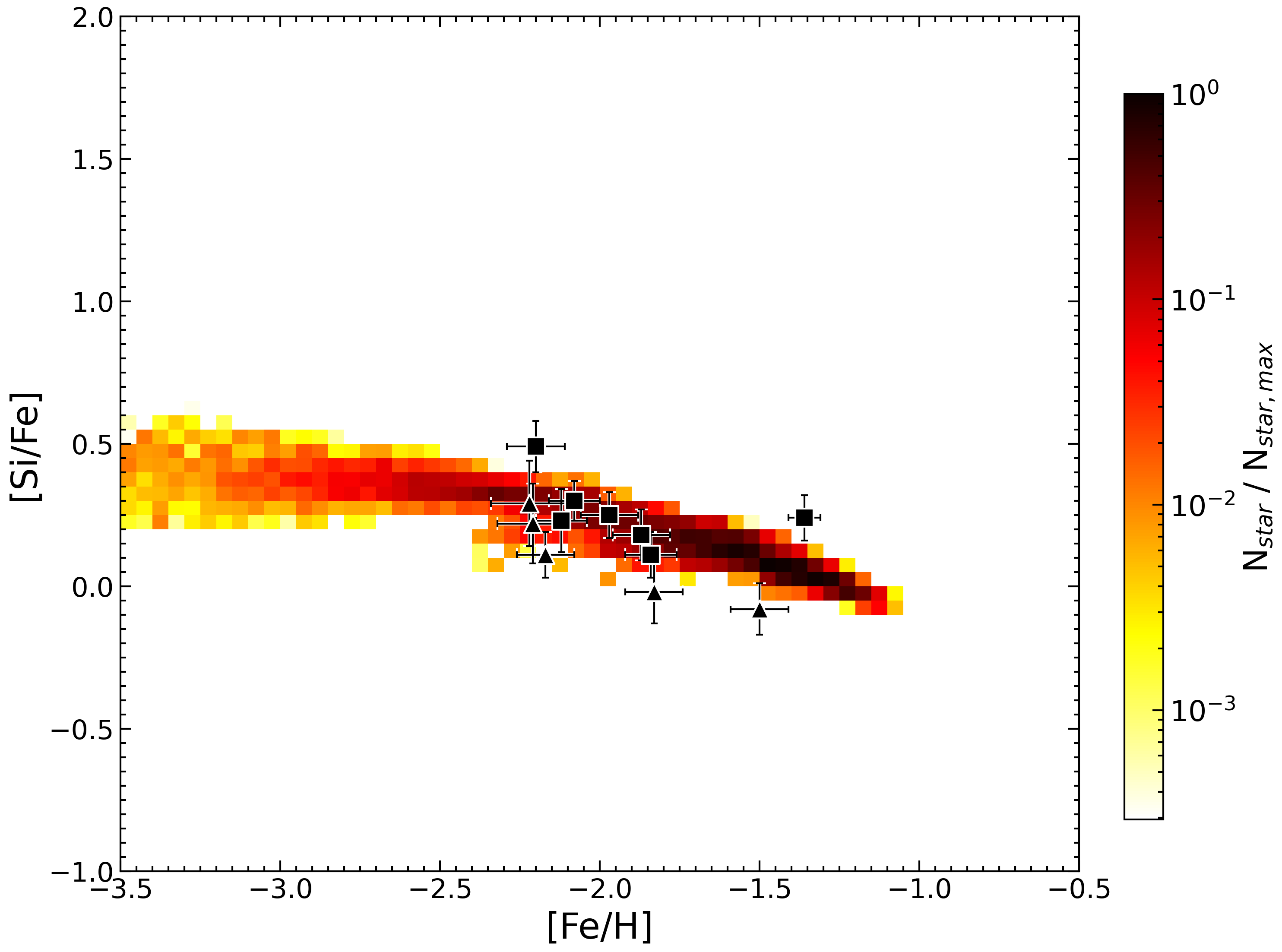}
        \caption{[O/Fe] (top panel) and [Si/Fe] (bottom panel) abundances as a function of [Fe/H] as predicted by our adopted stochastic chemical evolution model. The colormap displays the number of predicted long-lived stars by the model on a logarithmic scale. Black symbols are as in Fig. \ref{Eu_stochastic}.}
        \label{O_stochastic}
    \end{figure}

    Figure \ref{O_stochastic} shows the resulting [$\alpha$/Fe] vs. [Fe/H] abundance diagrams for O (top panel) and Si (bottom panel). Apart for the star MIKE-5, which shows a highly depleted [O/Fe] relative to other stars (and also detaching from the SMC observed trend at higher metallicity, see Figure \ref{alpha}), the prediction of the chemical evolution model both well reproduce the trend and the scatter of metal-poor SMC stars in O and Si. Therefore, the prediction of the model can be viewed as good tracer of the SMC early chemical evolution.

\subsection{Additional results on neutron-capture elements}
\label{aa:ncapt}

    Here, we provide the results obtained for additional run of the stochastic model, either i) decreasing the probability / fraction of system originating NSM in the mass range 8-50 \textit{M}$_\odot$ by a factor to $6 \times 10^{-3}$ (relative to the fiducial value of 0.08) or ii) changing the rotational velocity distribution by massive stars, favouring larger velocities relative to the fiducial distribution from \citet{Rizzuti21}\footnote{in \citet{Rizzuti21}, they adopt a Gaussian distribution centred on $\mu$ with dispersion $\sigma$ as the following: 
    \begin{align*}
        \mu = \left\{
        \begin{array}{ll}
        121.5 \ {\rm km\ s^{-1}} \hspace{3.15cm} {\rm for \ \ [Fe/H]}<-3\\[0.1cm]
        121.5 \cdot  e^{-2.324 \cdot ({\rm[Fe/H]}+3) } \ {\rm km\ s^{-1}} \hspace{1.13cm} {\rm for \ \ [Fe/H]}\geq-3 \\
        \end{array}, \right.  \\
        \sigma = \left\{
        \begin{array}{ll}
        114.2 \ {\rm km\ s^{-1}} \hspace{3.15cm} {\rm for \ \ [Fe/H]}<-3\\[0.1cm]
        114.2 - 58.5 \cdot  ({\rm[Fe/H]}+3) \ {\rm km\ s^{-1}} \hspace{0.41cm} {\rm for \ \ [Fe/H]}\geq-3 \\
        \end{array}. \right.
    \end{align*}}.

\begin{figure}
        \centering
        \includegraphics[width=0.475\textwidth]{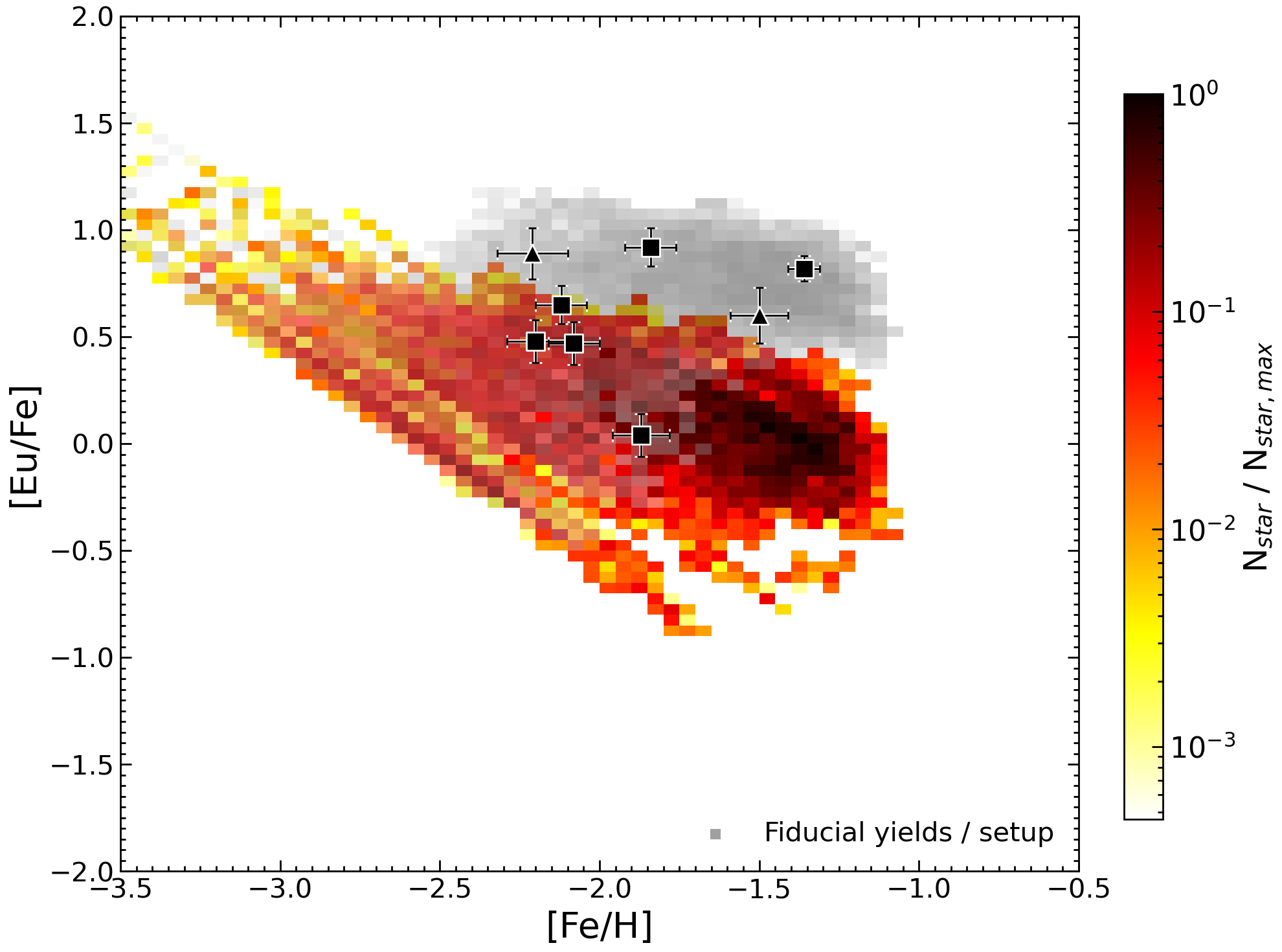}
        \caption{[Eu/Fe] abundances as a function of [Fe/H] as predicted by our adopted stochastic chemical evolution model with a reduced fraction of systems originating NSM. The colormap displays the number of predicted long-lived stars by the model on a logarithmic scale. As a comparison, we plot in grayscale the result by the stochastic chemical evolution model using our fiducial setup. Black symbols are as in Fig. \ref{Eu_stochastic}.}
        \label{Eu_compare}
    \end{figure}

    [Eu/Fe] vs. [Fe/H] predictions for the model with reduced fraction of systems originating NSM are displayed in Figure \ref{Eu_compare} (with the usual colormap). For comparison, the Figure also shows the results of our fiducial model (grayscale). As pointed out in the main text, the reduction in $r$-process production by delayed sources leads to a deficiency of Eu relative to what observed in the upper envelope of the SMC data.
    
    \begin{figure}
        \centering
        \includegraphics[width=0.475\textwidth]{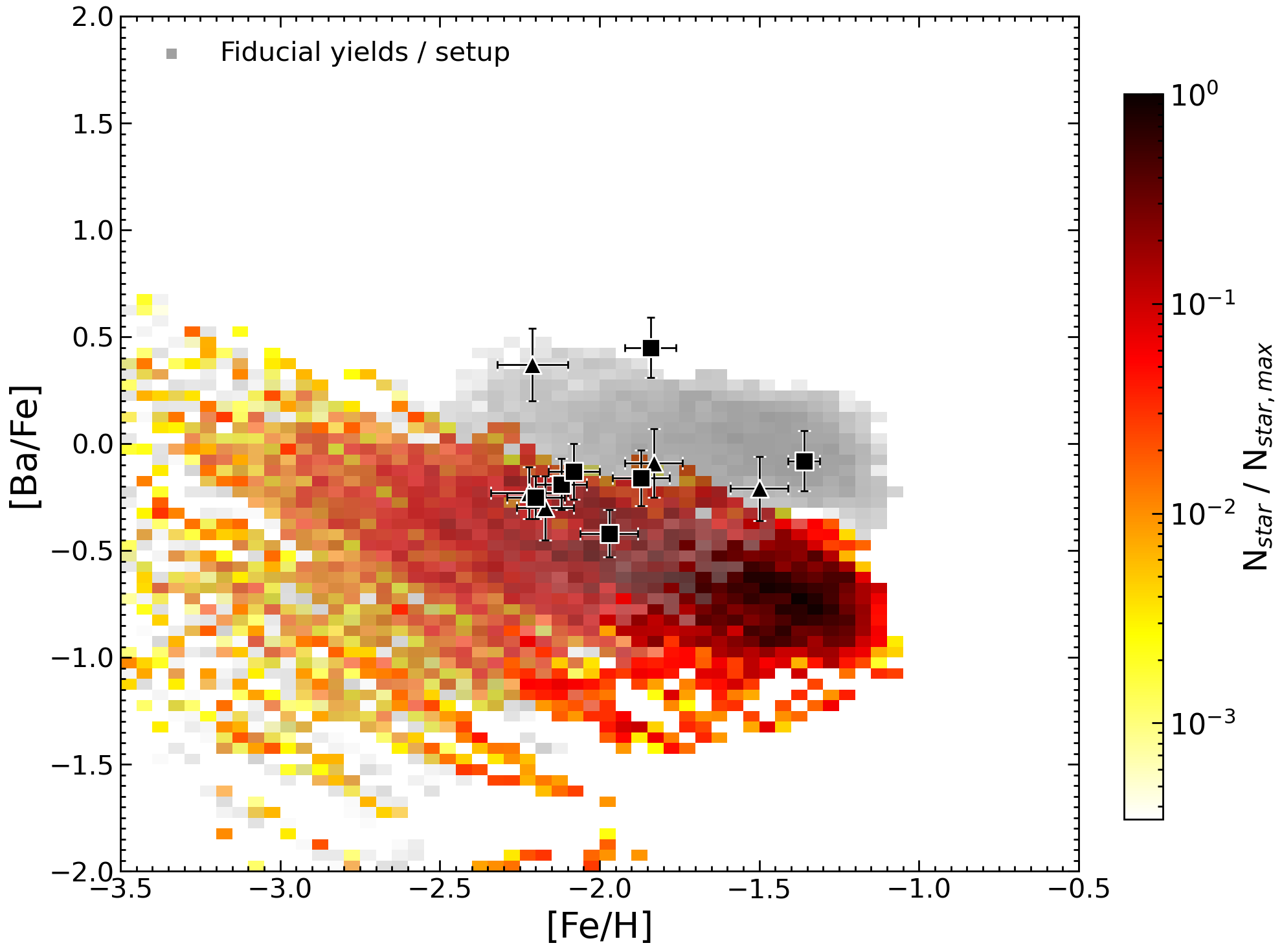}\\
        \includegraphics[width=0.475\textwidth]{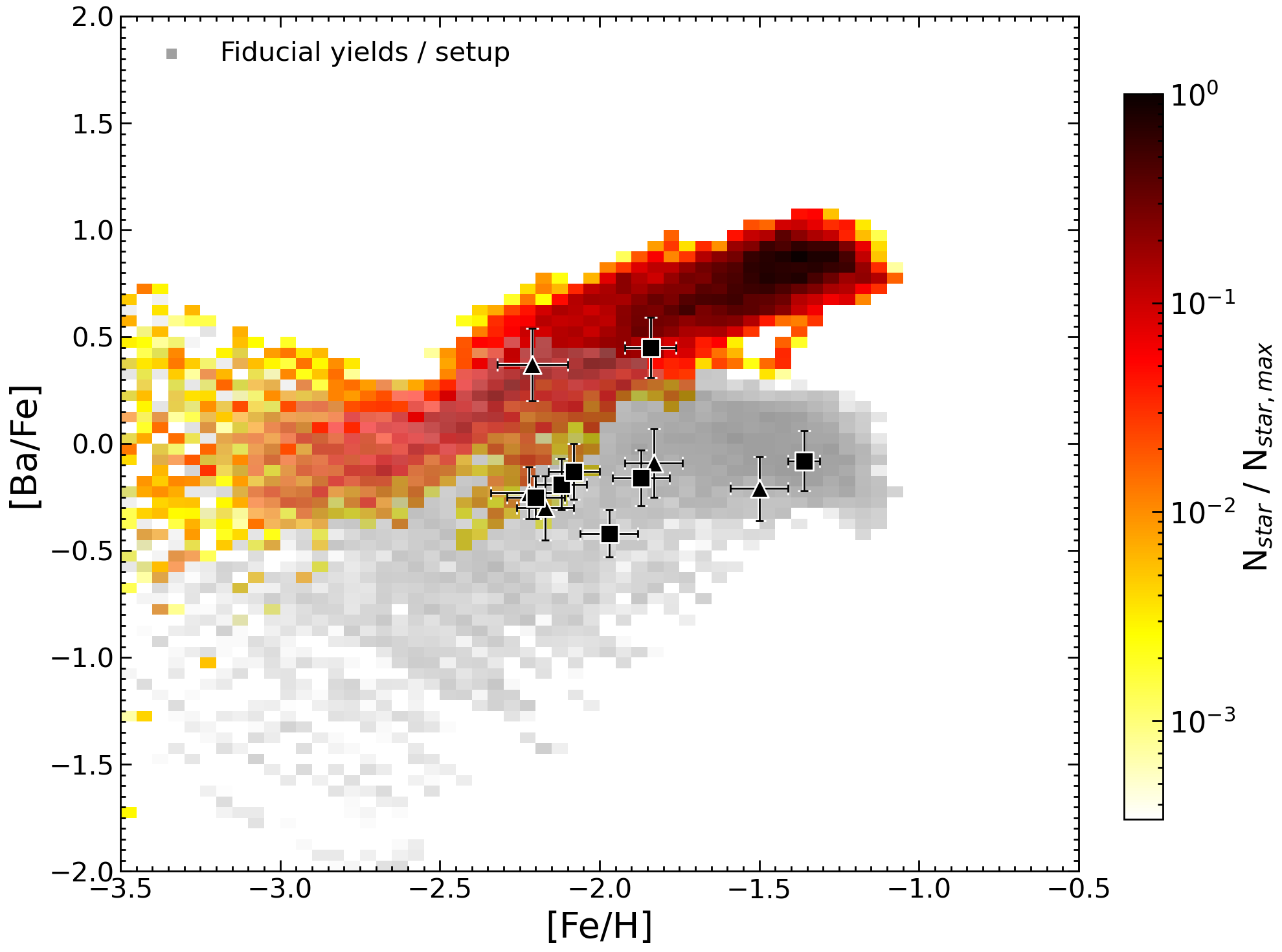}
        \caption{[Ba/Fe] abundances as a function of [Fe/H] as predicted by our adopted stochastic chemical evolution model with a reduced fraction of systems originating NSM (top panel) and rotational velocity distribution for massive stars favouring higher velocities (bottom panel). The colormap displays the number of predicted long-lived stars by the model on a logarithmic scale. As a comparison, we plot in grayscale the result by the stochastic chemical evolution model using our fiducial setup. Black symbols are as in Fig. \ref{Eu_stochastic}.}
        \label{Ba_sfrac}
    \end{figure}

    Such a reduction in $r$-process nucleosynthesis also affects the [Ba/Fe] vs. [Fe/H] predicted pattern. Indeed, Figure \ref{Ba_sfrac} top panel similarly shows that the model with reduced NSM fraction hardly reproduces a large fraction of stars with Ba abundances available, as expected due to the $r$-process driven production in the early galactic evolution (see also Figure \ref{Ba_stochastic}). On the other hand, Figure \ref{Ba_sfrac} bottom panel shows the [Ba/Fe] vs. [Fe/H] outcome for the model with different massive star rotational velocity distribution. In particular, for this model we adopt a uniform velocity distribution between 150 and 300 km s$^{-1}$, resulting in larger values than in our fiducial distribution, especially at large metallicity. Here the comparison with the fiducial model (in grayscale) shows a diverging behaviour for the two starting from [Fe/H]$\sim-2.5$ dex, with the new model showing a steady increase of [Ba/Fe] with metallicity, reaching values $\gtrsim 0.5$ dex at [Fe/H]$\sim-1.5$ dex. While the new model may help in the reproduction of two most Ba rich stars between $-2.5<$[Fe/H]/dex$<-2$, such a steady increase is not followed in the stars of our sample, suggesting that large rotational velocities are disfavoured at higher metallicities, as in the case of our fiducial distribution (see also \citealt{Prantzos18,Romano19}). 

\end{document}